\documentclass[twocolumn]{aastex63}

\usepackage{amsmath}
\usepackage{color,soul}
\usepackage{natbib}
\usepackage{float}

\newcommand{\kms}{\ensuremath{\rm km\,s^{-1}}}
\defcitealias{conroy_early-type_2014}{C14}
\defcitealias{beverage_elemental_2021}{B21}
\usepackage{lineno}
\shorttitle{Elemental Abundance Patterns of \texorpdfstring{$z\sim0.7$}{z~0.7} Quiescent Galaxies}
\shortauthors{Aliza G. Beverage}

\begin{document}



\title{From Carbon to Cobalt: Chemical compositions and ages of \texorpdfstring{$z\sim0.7$}{z~0.7} quiescent galaxies}

\correspondingauthor{Aliza G. Beverage}
\email{abeverage@berkeley.edu}

\author[0000-0002-9861-4515]{Aliza G. Beverage}
\affiliation{Department of Astronomy, University of California, Berkeley, CA 94720, USA}
\author[0000-0002-7613-9872]{Mariska Kriek}
\affiliation{Leiden Observatory, Leiden University, P.O. Box 9513, 2300 RA Leiden, The Netherlands}
\author[0000-0002-1590-8551]{Charlie Conroy}
\affiliation{Center for Astrophysics \textbar\ Harvard \& Smithsonian, Cambridge, MA, 02138, USA}
\author[0000-0002-7393-3595]{Nathan R. Sandford}
\affiliation{Department of Astronomy, University of California, Berkeley, CA 94720, USA}
\author[0000-0001-5063-8254]{Rachel Bezanson}
\affiliation{Department of Physics and Astronomy, University of Pittsburgh, Pittsburgh, PA 15260, USA}
\author[0000-0002-8871-3026]{Marijn Franx}
\affiliation{Leiden Observatory, Leiden University, P.O. Box 9513, 2300 RA Leiden, The Netherlands}
\author[0000-0002-5027-0135]{Arjen van der Wel}
\affiliation{Sterrenkundig Observatorium, Universiteit Gent, Krijgslaan 281 S9, B-9000 Gent, Belgium}
\author[0000-0002-6442-6030]{Daniel R. Weisz}
\affiliation{Department of Astronomy, University of California, Berkeley, CA 94720, USA}




\begin{abstract}
    We present elemental abundance patterns (C, N, Mg, Si, Ca, Ti, V, Cr, Fe, Co, and Ni) for a population of 135 massive quiescent galaxies at $z\sim0.7$ with ultra-deep rest-frame optical spectroscopy drawn from the LEGA-C survey. We derive average ages and elemental abundances in four bins of stellar velocity dispersion ($\sigma_v$) ranging from 150~km\,s$^{-1}$ to 250~km\,s$^{-1}$ using a full-spectrum hierarchical Bayesian model. The resulting elemental abundance measurements are precise to 0.05\,dex. The majority of elements, as well as the total metallicity and stellar age, show a positive correlation with $\sigma_v$. Thus, the highest dispersion galaxies formed the earliest and are the most metal-rich. We find only mild or non-significant trends between [X/Fe] and $\sigma_v$, suggesting that the average star-formation timescale does not strongly depend on velocity dispersion. To first order, the abundance patterns of the $z\sim0.7$ quiescent galaxies are strikingly similar to those at $z\sim0$. However, at the lowest velocity dispersions the $z\sim0.7$ galaxies have slightly enhanced N, Mg, Ti, and Ni abundance ratios and earlier formation redshifts than their $z\sim0$ counterparts. Thus, while the higher-mass quiescent galaxy population shows little evolution, the low-mass quiescent galaxies population has grown significantly over the past six billion years. Finally, the abundance patterns of both $z\sim0$ and $z\sim0.7$ quiescent galaxies differ considerably from theoretical prediction based on a chemical evolution model, indicating that our understanding of the enrichment histories of these galaxies is still very limited.
\end{abstract}


\section{Introduction}
\label{sec:intro}
The chemical composition of a galaxy reflects the complex interplay of many factors such as inflow rate of gas, outflow rate of metal-rich gas, star formation efficiency, and merger history. Whereas the composition of interstellar \textit{gas} in a galaxy provides an instantaneous snapshot of its metal-content, the composition of its \textit{stars} encodes the integrated enrichment of the gas over its entire star-formation and assembly history. Thus, the chemical makeup of the stars in a galaxy is a powerful probe of their star-forming and assembly histories \citep[e.g.][and references therein]{maiolino_re_2019}. Stellar chemical abundances are a particularly effective probe in quiescent galaxies, which have prominent stellar absorption features.

Stellar population modeling and elemental abundances at $z\sim0$ have yielded crucial insights into the formation and evolution of the massive quiescent galaxy population. One important finding is that the most massive quiescent galaxies in the local universe are also the most metal-rich \citep[e.g.][]{trager_stellar_2000, gallazzi_ages_2005, thomas_epochs_2005}. This finding is thought to reflect the strength of the gravitational potential of the galaxy; supernovae explosions and stellar winds in galaxies with deeper potential wells are less effective at removing metal-rich gas \citep[e.g.][]{larson_effects_1974, dekel_origin_1986, tremonti_origin_2004}. Massive quiescent galaxies are also found to be the oldest and most enriched in $\alpha$-elements as traced by the ratio [$\alpha$/Fe] \citep[e.g.][]{thomas_epochs_2005, conroy_early-type_2014, mcdermid_atlas3d_2015}. [$\alpha$/Fe] reflects the relative enrichment by prompt core-collapse and delayed Type Ia supernovae and thus directly probes the star-formation timescale of a galaxy \citep{matteucci_abundance_1994,trager_stellar_2000}. These results imply that the most massive galaxies in the local universe formed the bulk of their stars earliest and over the shortest timescales. Finally, detailed elemental abundance patterns have helped reveal the nucleosynthetic origins of various elements in massive galaxies. For example, there are indications that Fe-peak element Co may have significant contribution from core-collapse \citep{conroy_early-type_2014}, while  the $\alpha$-elements Ca and Ti may have significant contribution from Type Ia supernovae \citep[e.g.][]{saglia_puzzlingly_2002, thomas_stellar_2003, graves_ages_2007, johansson_chemical_2012}.

Whereas unresolved archaeological studies of $z\sim0$ galaxies have revealed a lot about the formation of quiescent galaxies, they also have their drawbacks. First, the abundances of stars in nearby galaxies only tell us about the star-formation histories of \textit{all} the stars currently in the galaxy, including younger and metal-poor stars accreted during late-time mergers. And second, inferring ages and star-formation histories becomes increasingly more difficult for old stellar populations. To overcome these challenges we must study the chemical compositions of galaxies at earlier epochs. Unfortunately, detailed stellar population modeling requires ultra-deep spectroscopy of the stellar continuum emission. As we push to higher redshift, the required signal-to-noise ratio (S/N) becomes increasingly expensive to reach. For this reason, the few existing studies of chemical abundances and stellar population properties of massive galaxies beyond $z\sim0$ have small sample sizes and only focus on a total metallicity, age, and sometimes an $\alpha$-abundance \citep[e.g.][]{gallazzi_charting_2014, kriek_massive_2016, leethochawalit_evolution_2019, kriek_stellar_2019, jafariyazani_resolved_2020, beverage_elemental_2021, carnall_stellar_2022}. Only \citet{choi_assembly_2014} measured additional abundances (C, N, and Ca) by stacking quiescent SDSS galaxies out to $z\sim0.7$.

The Large Early Galaxy Astrophysics Census (LEGA-C) survey is a major step forward in characterizing the $z\sim0.7$ massive galaxy population \citep{van_der_wel_vlt_2016, straatman_large_2018,van_der_wel_large_2021}. With ultra-deep continuum spectra for thousands of massive galaxies at $0.6<z<1.0$ it is now finally possible to measure detailed elemental abundances and stellar population properties from absorption line spectroscopy. Here we apply full-spectrum stellar population modeling to the LEGA-C sample, and present detailed abundance patterns of quiescent galaxies at $z\sim0.7$. We employ a hierarchical Bayesian method to measure the average abundances for galaxies in bins of velocity dispersion. Additionally, we revisit the $z\sim0$ SDSS stacks from \citet{conroy_early-type_2014} and re-fit them updated stellar population synthesis (SPS) models. We compare the updated $z\sim0$ abundance patterns with what we find at $z\sim0.7$ to probe the evolution of massive quiescent galaxies over the last six billion years.

The paper is organized as follows. In Section \ref{sec:data} we discuss the data and sample selection, in Section \ref{sec:fitting} we describe the fitting methods, and in Section \ref{sec:results} we present the abundance results. A discussion of our results and a comparison to SDSS are presented in Section \ref{sec:discussion}, along with a summary in Section \ref{sec:conclusion}. Throughout this work we assume a flat $\Lambda$CDM cosmology with $\Omega_{\rm m}= 0.29$ and $H_{\rm 0} = 69.3$\;km\,s$^{-1}$\,Mpc$^{-1}$ and Solar abundances of \citet{asplund_chemical_2009}, such that
Z$_\odot = 0.0142.$

\section{Data}
\label{sec:data}
\subsection{LEGA-C}
We select our sample from the third data release of the Large Early Galaxy Census (LEGA-C) Public Spectroscopic Survey \citep{van_der_wel_vlt_2016, straatman_large_2018, van_der_wel_large_2021}. LEGA-C is a deep spectroscopic survey targeting 3528 galaxies at intermediate redshifts ($0.6\lesssim z \lesssim 1.0$) in the COSMOS footprint \citep{scoville_cosmic_2007}. The galaxies were selected from the public UltraVISTA catalog \citep{muzzin_public_2013} using a redshift-dependent $K_s$ magnitude limit, which range from $K_s=21.0-20.4$. The 20-hr spectra were collected using VIMOS on the VLT, and have an average continuum S/N of 20\AA$^{-1}$. For further details on survey design, we refer to \cite{straatman_large_2018}, and for details regarding the specifics of DR3 we refer to \cite{van_der_wel_large_2021}.

\begin{figure*}
    \centering
    \includegraphics[width=\textwidth]{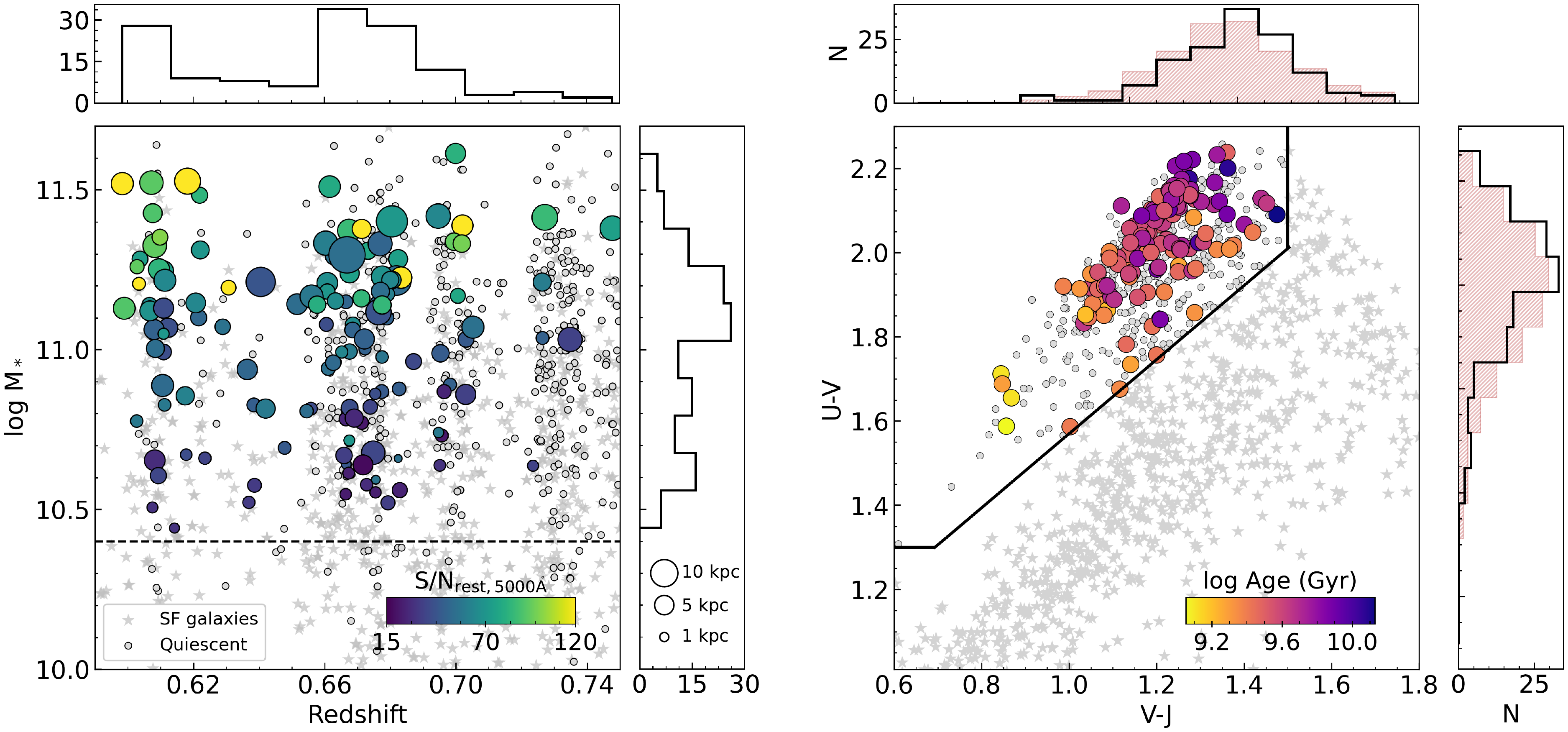}
    \caption{\textit{Left:} The selected sample in redshift vs stellar mass space, colored by S/N at rest-frame 5000\,\AA. The sizes of each point correspond to the physical half-light radius of the object, with the smallest and largest circles representing 1\,kpc and 10\,kpc, respectively. At higher redshift, fewer quiescent galaxies are included in the sample, as they do not reach the required wavelength range of 5350\AA. The dashed line shows the 90\% mass-completeness limit (log\,M/M$_\odot$ = 10.4). \textit{Right:} The selected sample in \textit{UVJ} space, colored by best-fit stellar ages from \texttt{alf}. For comparison, both panels show the full LEGA-C dataset for the selected redshift range ($0.6<z<0.75$). The small stars represent star-forming galaxies, and small circles are quiescent, according to their position on the UVJ diagram \citep{muzzin_evolution_2013}. The black histograms show the distribution of the selected sample along each axis. For the $UVJ$ panel we also include distributions of the quiescent LEGA-C dataset in red, normalized to the area of the black histograms.}
    \label{fig:sel}
\end{figure*}

\subsubsection{Stellar masses and rest-frame colors}
We measure stellar masses and rest-frame colors using the multi-wavelength UltraVISTA photometric catalog \citep{muzzin_public_2013}. We determine rest-frame \textit{UVJ} colors using EAZY \citep{brammer_eazy_2008} and measure stellar masses with the \texttt{FAST} fitting code \citep{kriek_ultra-deep_2009}, fixing the input redshifts to the spectroscopic redshifts provided in the LEGA-C catalog. In the \texttt{FAST} fitting we use Flexible Stellar Population Synthesis \citep[\texttt{FSPS}]{conroy_propagation_2009} templates, assuming a delayed exponentially declining star formation history, the \citet{chabrier_galactic_2003} initial mass function (IMF), the \citet{kriek_dust_2013} dust attenuation law, and solar metallicity. The impact of using solar metallicity and solar-scaled abundance pattern templates on stellar mass estimates is further explored in Beverage et al. (\textit{in prep}). Finally, we correct the stellar masses such that they are consistent with the best-fit S\'ersic profiles by multiplying the stellar masses by the ratio of the S\'ersic model flux in F814W and the interpolated F814W flux from the photometric catalog \citep[e.g.][]{taylor_masses_2010}. On average, this procedure increases the stellar mass by 4\%. We compare our corrected \texttt{FAST} stellar masses to those derived using \texttt{MAGPHYS} \citep{da_cunha_simple_2008, de_graaff_observed_2021} and find no systematic offset and a scatter that is consistent with the uncertainties on individual stellar mass measurement ($\sim0.1$\,dex). We refer to \citet{de_graaff_observed_2021} for a detailed comparison of the two methods.

\subsubsection{Sample selection}
We use the stellar masses and rest-frame colors to select a mass-complete quiescent galaxy sample from the LEGA-C catalog. We begin by requiring each galaxy spectrum to cover H$\beta$, Mgb, and at least two FeI features (rest-frame $4300$\,\AA$ < \lambda < 5340\,$\AA), which translates to a redshift limit of $z<0.75$. This wavelength regime is crucial for accurately modeling the Mg and Fe abundances. Next, we classify galaxies as quiescent and star-forming using the rest-frame $U-V$ and $V-J$ colors and the selection criteria from \citet{muzzin_evolution_2013}. After removing the star-forming galaxies, we  enforce a 95\% completeness limit of $\log\,\mathrm{M/M_\odot}=10.4$ corresponding to a S/N limit of 15\AA$^{-1}$ following the procedure outlined in \citet{beverage_elemental_2021}. The completeness limit matches what is found in \cite{van_der_wel_large_2021} at $z\sim0.65$. 

The above selection criteria produce a mass-complete sample of 135 quiescent galaxies. The right panel of Figure~\ref{fig:sel} shows the selected sample in \textit{UVJ} space, along with the full LEGA-C dataset at $0.6<z<0.75$. The left panel of Figure~\ref{fig:sel} shows the sample in redshift vs stellar mass space, colored by the S/N at rest-frame 5000\AA. The size of each point represents the physical half-light radius in kpc as measured in \textit{HST} ACS F814W (rest-frame $\sim5000$\,\AA) from the public LEGA-C catalog. Briefly, sizes were measured from 10'' cutouts by fitting a single-component S\'ersic profile to the public COSMOS imaging \citep{scoville_cosmic_2007} using  \texttt{galfit} \citep{peng_detailed_2010} \citep[for details refer to][]{van_der_wel_structural_2012, van_der_wel_vlt_2016}. The bias of our selection towards galaxies at lower redshifts is due to the imposed spectral range criterion.

\begin{figure*}
    \centering
    \includegraphics[width=\textwidth]{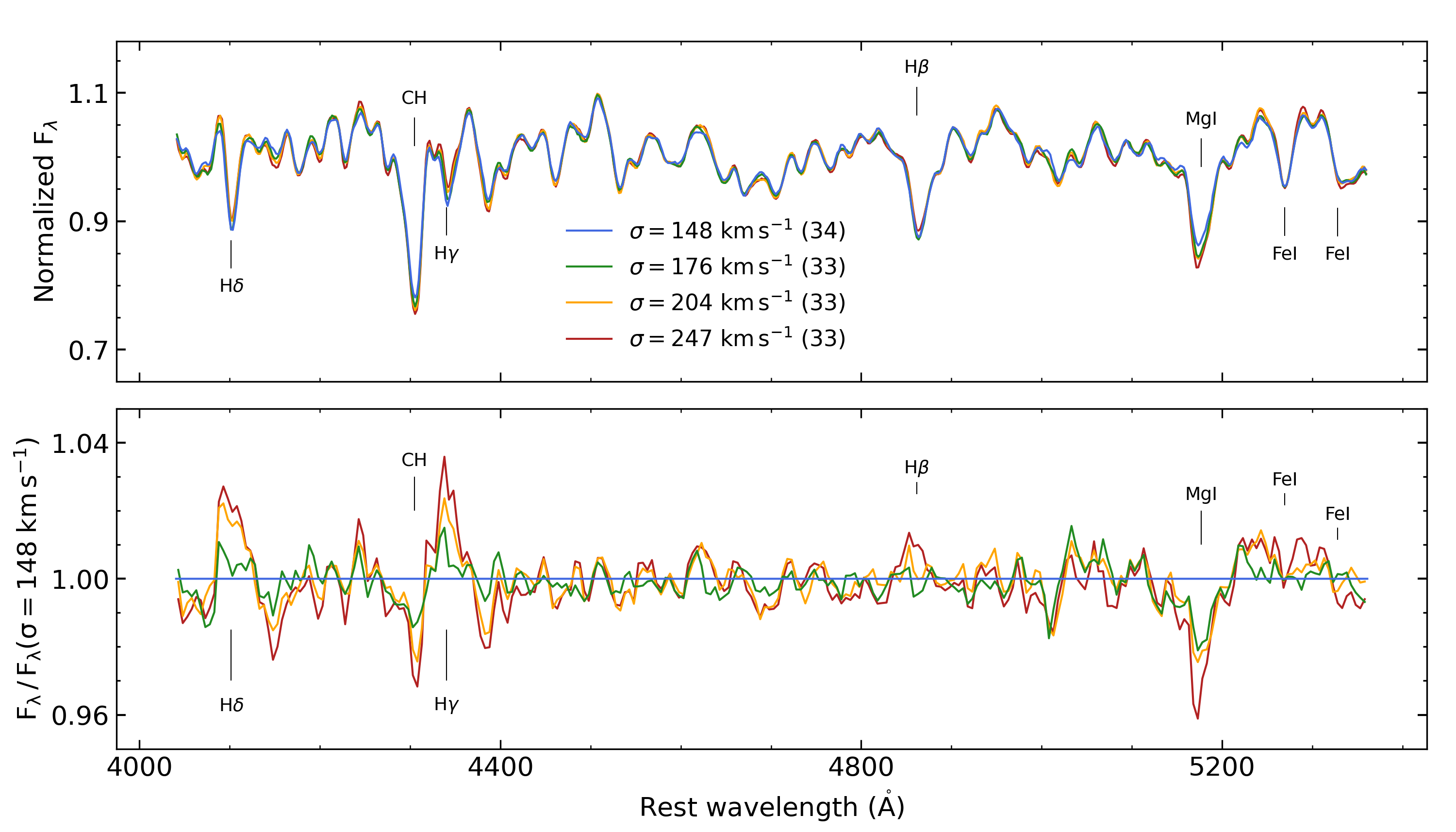}
    \caption{Continuum-normalized stacked LEGA-C spectra in bins of increasing velocity dispersion. Before stacking, each spectrum is smoothed to the same velocity dispersion (300\;km\,s$^{-1}$). The lower panel shows the ratio of the flux in each stack divided by the flux of the lowest velocity dispersion bin. The listed velocity dispersions in the legend are the median observed values of the individual galaxies in each bin. In parenthesis we list the number of individual spectra that went into each stack. These stacked spectra are not fit directly (see Section \ref{sec:hbm}), but demonstrate the high quality of the LEGA-C spectra. It is clear from both panels that galaxies with larger velocity dispersions have deeper metal features and shallower Balmer lines.}
    \label{fig:spectra}
\end{figure*}

To illustrate the quality of the LEGA-C spectra, we split the galaxies into four bins of velocity dispersion and stack their spectra. The velocity dispersions are taken from the LEGA-C DR3 catalog. In order to combine spectra with different continuum shapes and normalizations, we first divide each spectrum by an $n=7$ polynomial. We find that polynomials with degree $4<n<9$ successfully remove the broad continuum shape of the spectra while preserving their absorption features. Next, we smooth each spectrum to a common velocity broadening by convolving each spectrum with a Gaussian kernel to achieve an effective dispersion of 300\;km\,s$^{-1}$. The continuum-normalized and smoothed spectra are coadded by taking the median flux at each wavelength.

The stacked spectra are shown in the top panel of Figure~\ref{fig:spectra}. In the bottom panel we show the flux of each stack divided by the flux of the lowest velocity dispersion bin. While the differences between the stacked spectra are small, the combined S/N of the LEGA-C spectra reveal subtle trends; the Balmer lines become shallower with increasing velocity dispersion, while the MgI and CH features become deeper. By eye it is impossible to untangle whether these trends with velocity dispersion are due to underlying trends with age, metallicity, or both; we turn to full-spectrum hierarchical Bayesian modeling to quantify the age, metallicity, and individual abundances separately.

\subsection{SDSS comparison sample}
\label{sec:sdss_compare}
We use the early-type galaxies from SDSS of \citet[][C14]{conroy_early-type_2014} as a low-redshift comparison sample. \citetalias{conroy_early-type_2014} stacked thousands of massive early-type galaxies in bins of velocity dispersion and measured their abundance patterns. In this paper we use the same stacks but re-fit the spectra with updated stellar population models (see next section). We refer to \citetalias{conroy_early-type_2014} for details on their sample selection and stacking method. Briefly, the galaxies were selected from the SDSS Main Galaxy Survey Data Release 7 \citep{abazajian_seventh_2009} within the redshift interval $0.025<z<0.06$. Passive galaxies were identified by requiring no emission in H$\alpha$ or [OII]$\lambda$3727, and the sample was further restricted to galaxies that lie on the Fundamental Plane (FP) as defined by the central FP slice in \citet{graves_dissecting_2010}. The LEGA-C quiescent galaxies in the current study were instead selected based on their $UVJ$ colors only and not emission line fluxes, mainly because most of the LEGA-C spectra do not cover the H$\alpha$ and [OII]$\lambda$3727 lines. However, \citet{maseda_ubiquitous_2021} show that most quiescent galaxies in LEGA-C with [OII]$\lambda$3727 coverage indeed have detected nebular low-ionization emission most likely originating from evolved stars (i.e., post-AGB and blue horizontal branch stars) and not low-level star-formation. Furthermore, at low-redshift it has been shown that this emission is unlikely to originate from low-luminosity AGN \citep{yan_nature_2012}. 

The SDSS spectra were continuum-normalized and convolved to an effective velocity dispersion of 350 \kms\ before stacking. At this redshift, the SDSS fiber samples the inner $\sim0.5\,R_e$, whereas the LEGA-C slit on average samples slightly larger than $1\,R_e$ in the wavelength direction.  The slit still captures a fraction of light at large radii in the spatial direction, however due to optimal extraction weighing, the spectra are still dominated by the centers. We discuss the implications of this aperture difference in Section~\ref{sec:discussion}.

\section{Model Fitting}
\label{sec:fitting}
\subsection{Full-spectrum modeling with \texttt{alf}}
\label{sec:alf}

We derive stellar population parameters using the full-spectrum absorption line fitter code \texttt{alf} \citep{conroy_counting_2012, conroy_metal-rich_2018}. \texttt{alf} generates models by combining the metallicity-dependent MIST isochrones \citep{choi_mesa_2016} with the MILES and Extended-IRTF empirical stellar libraries \citep{sanchez-blazquez_medium-resolution_2006,villaume_extended_2017}. To adjust the stellar libraries for abundance variations in individual elements, \texttt{alf} uses metallicity- and age-dependent synthetic response functions to determine the fractional change in spectra due to the variation of each individual element. To match the continuum shape of the input spectrum to the \texttt{alf} models, a high-order polynomial with degree $n$, where $n\equiv(\lambda_{\rm max}-\lambda_{\rm min})/100$\,\AA, is fit to the ratio of data/model \citep[see][for details]{conroy_counting_2012}. This polynomial is then divided by the input spectrum. The posteriors are computed on the continuum-matched input spectrum using Markov-Chain Monte Carlo (MCMC) as implemented by the \texttt{emcee} package \citep{foreman-mackey_emcee_2013}.

\begin{figure*}
    \centering
    \includegraphics[width=\textwidth]{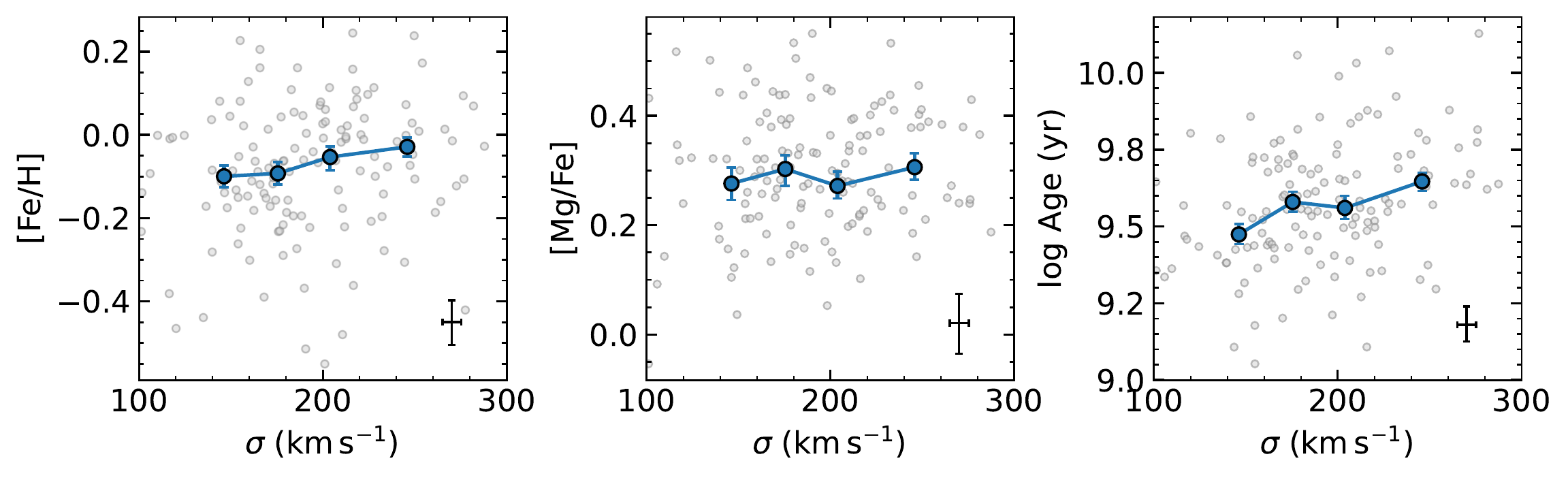}
    \caption{The [Fe/H], [Mg/Fe], and stellar age results for the individual galaxies in our sample (gray). Typical errorbars are shown in bottom right corner of each panel. As described in Section~\ref{sec:hbm}, the PDFs from the individual fits are used in the hierarchical Bayesian model to derive average abundances and ages in bins of $\sigma_v$. The blue points show the results of this modeling.}
    \label{fig:ind_res}
\end{figure*}

In this study, we configure \texttt{alf} with a \citet[]{kroupa_variation_2001} IMF and a single stellar population age. In total, we fit for 33 parameters with \texttt{alf}: recessional velocity, line-of-sight velocity dispersion (LOSVD) and corresponding hermite parameters $h_3$ and $h_4$, a single stellar population age, a scaling metallicity, 19 individual elemental abundances (Fe, C, N, O, Na, Mg, Si, Ca, Ti, K, V, Cr, Mn, Ni, Co, Sr, Ba, Eu), six emission line fluxes (Balmer lines, [OII], [OIII], [SII], [NI], and [NII]), and two nuisance parameters (a temperature shift applied to the fiducial isochrones and a ``jitter'' term that accounts for over- or under-estimated uncertainties on the data). fix the IMF to \citet[]{kroupa_variation_2001}. For the MCMC fits we use 1024 walkers with a 20,000-step burn-in and assume uniform priors on all model parameters. In Figure~\ref{fig:ind_res} we show the results from the individual fits for [Fe/H], [Mg/Fe] and stellar age. These results are used in the next section to derive the average abundances and stellar ages in bins of velocity dispersion. Example LEGA-C spectra with corresponding best-fit models can be found in \citetalias{beverage_elemental_2021}.

In Appendix~\ref{app:sim_stack} we test the robustness of our fitting by simulating LEGA-C spectra and attempting to retrieve the true values with \texttt{alf}. This test is used to determine which element can be confidently constrained. We only include parameters with reduced $\chi^2<2$ and uncertainties $<0.15$\,dex. For the LEGA-C spectra, that includes 11 out of the 19 elements (C, N, Mg, Si, Ca, Ti, V, Cr, Fe, Co, and Ni).

\begin{figure}
    \centering
    \includegraphics[width=\columnwidth]{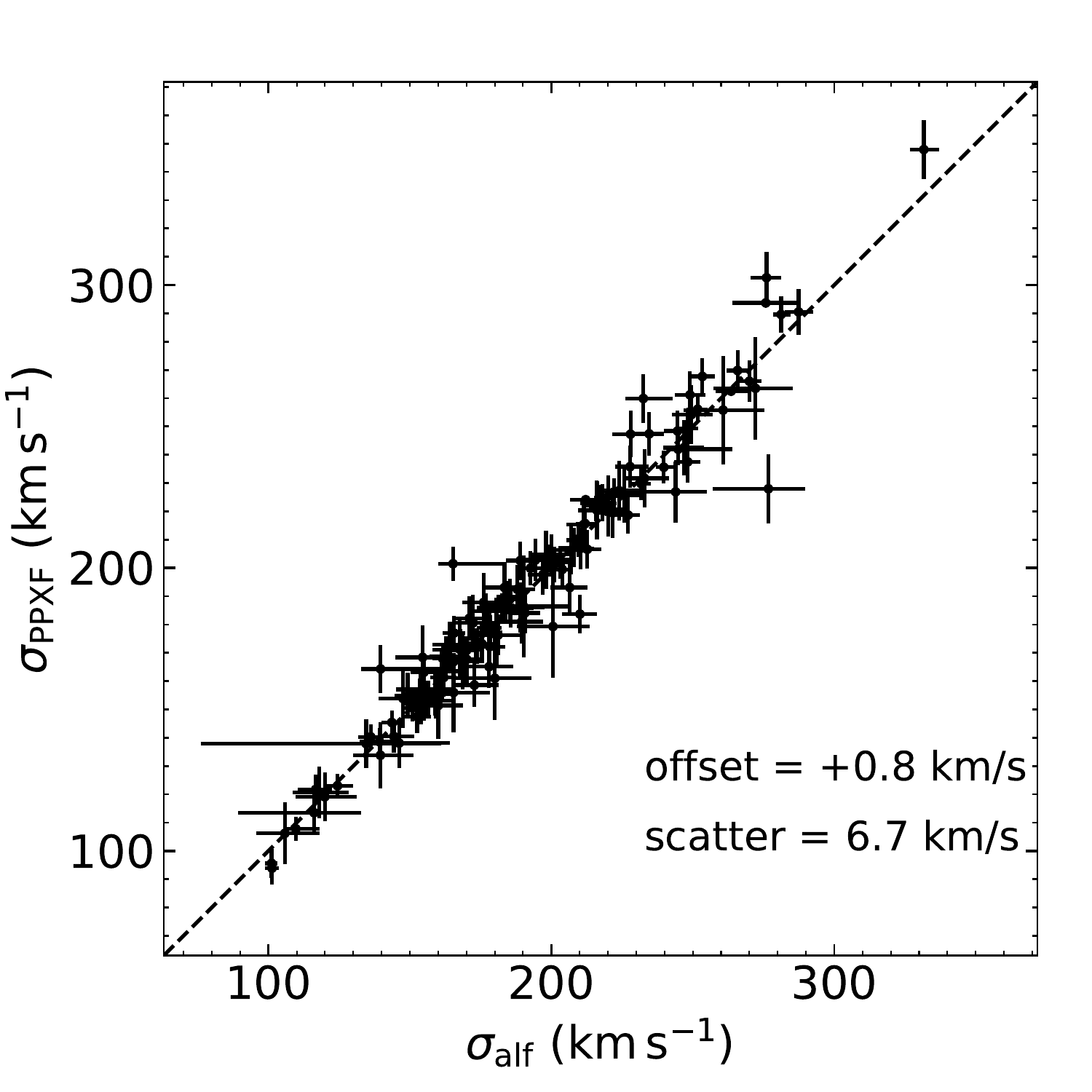}
    \caption{Comparison between velocity dispersions from \texttt{alf} (this study) and from \texttt{PPXF} (reported by the LEGA-C team). Both methods use full-spectrum modeling, but with different underlying stellar libraries. The two methods are in good agreement.}
    \label{fig:compare_sigmas}
\end{figure}

In Figure~\ref{fig:compare_sigmas} we compare the velocity dispersions derived using \texttt{alf} with the publicly-available LEGA-C values measured using \texttt{PPXF} \citep{cappellari_parametric_2004, cappellari_improving_2017, bezanson_spatially_2018, cappellari_full_2022}. Both methods use full-spectrum fitting, but differ in their underlying model assumptions and stellar libraries. Nonetheless, the velocity dispersions are in very good agreement, with the offset nearly zero, and the scatter of the data consistent with the errors on the individual measurements.

As an independent check of our stellar age measurements we show the selected sample in $UVJ$ space colored by the best-fit stellar ages from \texttt{alf}. It is clear that the oldest \texttt{alf} ages correspond to the redder $UVJ$ colors and thus older ages \citep[see][]{belli_mosfire_2019}. This result is reassuring given that the $UVJ$ ages -- measured using the continuum shape -- are independent from the \texttt{alf} ages, which rely solely on absorption features.

\citetalias{conroy_early-type_2014} also use \texttt{alf} to measure the stellar population parameters and elemental abundances of the stacks of massive early-type galaxies. However, since the publication of \citetalias{conroy_early-type_2014} several ingredients in the \texttt{alf} models have seen improvements, such as metallicity- and age-dependent response functions, along with updated isochrones and empirical spectral libraries \citep{villaume_extended_2017,conroy_metal-rich_2018}. Thus, we re-fit the stacked SDSS spectra with the improved \texttt{alf} models using identical settings to the LEGA-C spectra. We test various \texttt{alf} settings to ensure the SDSS results are not affected by fitting decisions. Our tests include fitting with a hot star component, setting the IMF as a free parameter, and adding a young stellar population component. The abundance results are affected by only $<0.05\,$dex, however the addition of the hot star component results in unrealistic ages. Thus, we turn off the hot star component for both LEGA-C and SDSS fitting.

SDSS covers a much redder wavelength range than LEGA-C and thus we also investigate the impact of  different spectral fitting regions on the \texttt{alf} results. When we limit the SDSS spectral range ($\rm{0.38\mu m - 0.64\mu m}$ and $\rm{0.80\mu m-0.88\mu m}$) to match LEGA-C ($\rm{0.38\mu m - 0.54\mu m}$) we find that the results are consistent but the uncertainties on the individual measurements increase as expected.

We also compare our LEGA-C abundance results to those from the previous data release \citep[DR2][]{van_der_wel_vlt_2016} presented in \citetalias{beverage_elemental_2021}. \citetalias{beverage_elemental_2021} measured individual Mg- and Fe-abundances and stellar population ages using the same \texttt{alf} setup as described here. However, since DR2, the LEGA-C catalog has nearly doubled in size and the spectrum reduction pipelines have since been updated \citep{van_der_wel_large_2021}. We compare the 65 overlapping galaxies in the \citetalias{beverage_elemental_2021} sample and the sample presented in this paper. Interestingly, we find $\sim0.2\,$dex difference in the absolute abundance measurements [Fe/H] and [Mg/H], while the [Mg/Fe] and age measurements are consistent. Upon closer inspection, the DR3 spectra indeed have deeper absorption features, as measured by their equivalent widths. This difference is likely due to improvements in the background subtraction and object extraction algorithms between DR2 and DR3. Further testing of systematic uncertainties on absolute abundances is required, and thus here we restrict the discussion between the $z\sim0.7$ and $z\sim0$ samples solely to the abundances ratios [X/Fe]. Nonetheless, we caution that this type of work can be extremely sensitive to background subtraction.

\subsection{Average ages and abundances of \texorpdfstring{$z\sim0.7$}{z~0.7} galaxies from hierarchical Bayesian modeling}
\label{sec:hbm}

In the previous section we derived stellar population properties and elemental abundances for individual LEGA-C galaxies. To assess the abundances of many more elements than possible for an individual galaxy, we derive averages using hierarchical Bayesian modeling. This approach has several advantages compared to fitting a stacked spectrum as shown in Appendix~\ref{app:sim_stack}. First, we do not have to smooth the spectra to a common velocity dispersion; smoothing introduces correlated noise and smooths out the absorption features. And second, we do not have to interpolate the spectra or remove the continuum by fitting polynomials, both of which introduce more correlated noise and may result in signal that is not real.

To overcome these concerns, we turn to hierarchical Bayesian modeling. In a hierarchical framework, models can be defined to have multiple ``levels," wherein the first level describes individual observations and the second level describes how the individual observations are distributed as a group. The individual- and population- level parameters can be evaluated simultaneously using Bayes' Theorem,

\begin{equation}
    \label{eq:bayes}
    P(\theta|X) = \frac{P(X|\theta)\cdot P(\theta)}{P(X)}
\end{equation}

\noindent where $\theta$ are the set of parameters describing the model and $X$ is the data. The left-hand side is the probability of the model given the data, also known as the posterior. The first term on the right-hand side is the likelihood, while the second term is the prior on the model parameters. The denominator is a normalizing constant, known as the ``evidence,'' which is often ignored. For hierarchical models, Bayes' Theorem can be re-written,

\begin{equation}
    \begin{aligned}
    \label{eq:bayes_theorem}
    P(\theta,\alpha|X) &= \frac{P(X|\theta, \alpha)\cdot P(\theta)\cdot P(\alpha)}{P(X)}\\
    &= \frac{P(X|\theta)\cdot P(\theta|\alpha)\cdot P(\theta)\cdot P(\alpha)}{P(X)}.
    \end{aligned}
\end{equation}

\noindent where $\theta$ is the set of parameters in the first (individual) level, and $\alpha$ describes how these parameters are distributed globally (population-level). The posteriors on both $\theta$ and $\alpha$ can be evaluated by integrating Equation \ref{eq:bayes_theorem}. In what follows, we describe our hierarchical model and how it is implemented.

\subsubsection{The Hierarchical Model}
We use a two-level model that directly mimics the results we would expect from stacking the LEGA-C spectra in bins of velocity dispersion. The first level of our hierarchical model is the full-spectrum population synthesis models in \texttt{alf}, described by various stellar population parameters. The second level describes how these parameters are distributed in the \textit{population} of galaxies ($\alpha$). We assume that they are distributed normally $\mathcal{N}(\mu_{\theta_i},\, \sigma_{\theta_i}^2)$, with a mean of $\mu_{\theta_i}$ and an intrinsic spread of $\sigma_{\theta_i}$. In this study we are primarily interested in $\mu_{\theta_i}$, and as such the population-level model choice has little impact on our conclusions. However, with a larger sample, it may be interesting to investigate trends in $\sigma_{\theta_i}$. Furthermore, future studies with larger samples can use a model selection criteria (e.g., Bayesian Information Criterion) to select the optimal population-level model and provide insight into the true distribution of stellar population parameters.


While the scope of this paper is limited to a normal distribution population model, we emphasize that this method can easily handle something more complicated. For example, an interesting extension would be to use a population-level model that asserts all galaxies follow a linear mass-metallicity relation. 

\begin{figure*}
    \centering
    \includegraphics[width=0.9\textwidth]{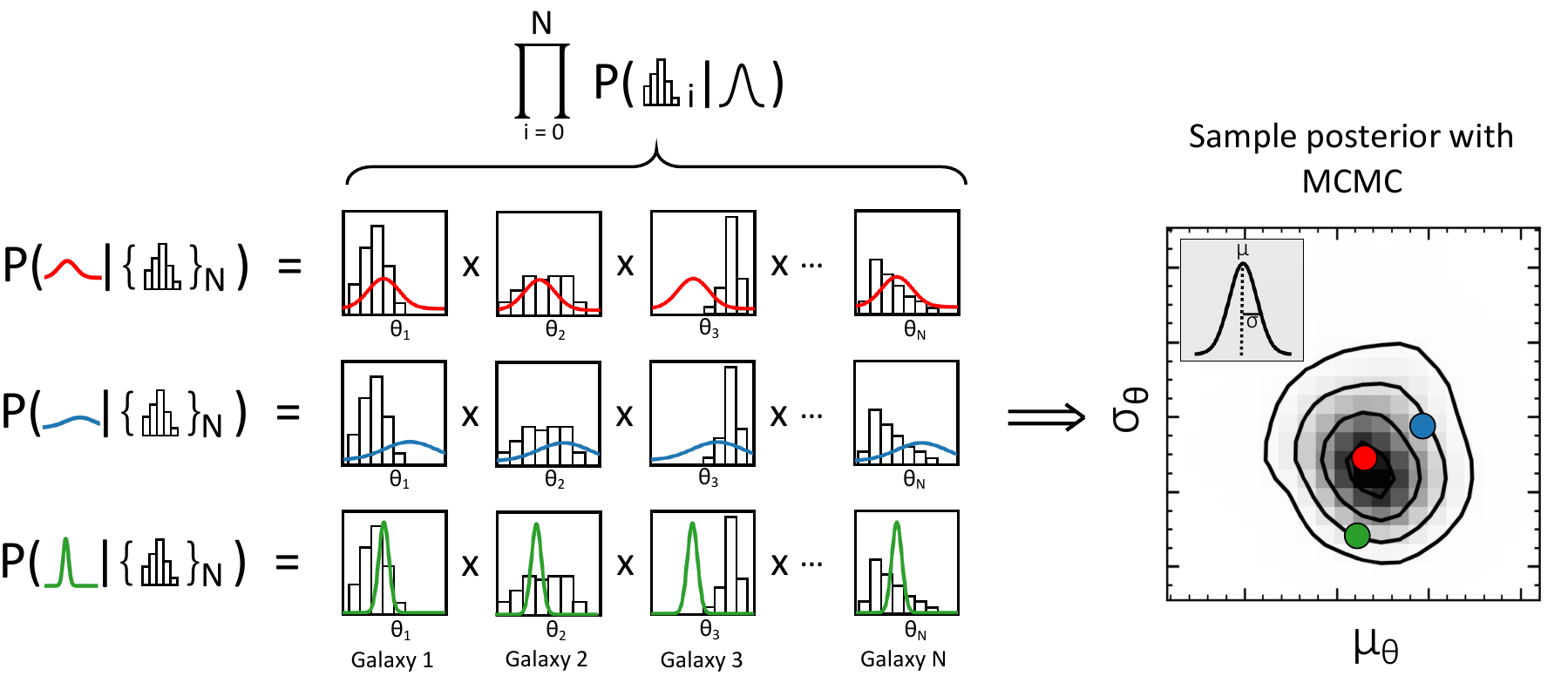}
    \caption{Illustration of our hierarchical Bayesian modeling implementation. An example posterior for the population-level mean $\mu_\theta$ and spread $\sigma_\theta$ for parameter $\theta$ is shown in the right panel. On the left we show how to evaluate the likelihood for three different population distributions (one for each row). Instead of modeling the individual- and population-level parameters simultaneously, we evaluate the population-level distribution using the results from individual full-spectrum fits. The likelihood function for the population model is defined by calculating the probability of each individual posterior (histograms in each column) given the model (normal distributions in each row) and taking the product of these probabilities over all N galaxies. The posteriors of $\mu_\theta$ and $\sigma_\theta$ are computed by sampling this likelihood function using MCMC.}
    \label{fig:hbm_demo}
\end{figure*}

\subsubsection{Implementation}

In hierarchical Bayesian modeling, typically the individual- and population-level parameters are evaluated simultaneously. However, in the case presented here, we use the posteriors from individual full-spectrum fits (Section \ref{sec:alf}) to evaluate the population model. The MCMC posterior chains from the individual full-spectrum fits $P(\theta|X)$ can be expressed by Eq.~\ref{eq:bayes}. Substituting this expression into Eq. \ref{eq:bayes_theorem}, we arrive at an expression for $P(\theta,\alpha|X)$, or the posterior on the population-level model,

\begin{equation}
    \label{eq:chian_bayes}
    P(\theta,\alpha|X) = P(\theta|X)\cdot P(\theta|\alpha)\cdot P(\alpha)
\end{equation}

\noindent where the first term on the left-hand side is the set of posteriors from the individual fits, the second term describes the population-level model, which in our case is $P(\theta|\alpha)\sim \mathcal{N}(\mu_{\theta_i},\, \sigma_{\theta_i}^2)$, and the final term is the prior on the population-level parameters $\alpha = (\mu_{\theta_i},\,\sigma_{\theta_i})$, which we assume to be uniform, i.e. $P(\alpha) \sim \mathcal{U}$. The above equation can be re-written in integral form,

\begin{equation}
    \label{eq:mcmc}
    P(\theta,\alpha|X) =  \prod_{n=1}^{N}\int P(\theta_n|X_n)\cdot P(\theta_n|\alpha)\,d\theta.
\end{equation}

These integrals can be numerically integrated with MCMC sampling.

In Figure~\ref{fig:hbm_demo} we illustrate the implementation of our hierarchical model. Take for example $\theta=\rm{[Fe/H]}$. As described in Section \ref{sec:alf} the [Fe/H] posteriors for each individual galaxy have already been derived. We use these posteriors to determine the mean ($\mu$) and intrinsic scatter ($\sigma$) of Fe-abundances for each velocity dispersion bin. In the left three rows of Figure~\ref{fig:hbm_demo} we show how to calculate the likelihood of three example population-level models. Essentially, we take the [Fe/H] posteriors for each galaxy in the bin (histograms in the middle columns) and calculate the probability of those posteriors given the particular model. Then to get the likelihood, we take the product of these probabilities over all $N$ galaxies in that bin. With enough MCMC samples it is eventually possible to populate a posterior for $\mu$ and $\sigma$ such as the one in the right panel.

We use the hierarchical Bayesian model to derive average stellar population properties and elemental abundances of the LEGA-C galaxies in four bins of velocity dispersion. For consistency with our abundance results, the binning is done using velocity dispersions from the individual \texttt{alf} fits instead of the \texttt{PPXF} measurements in the LEGA-C public catalog, although the two measurements are in good agreement (see Figure~\ref{fig:compare_sigmas}). This method is prohibitively expensive at the scale of thousands of spectra included in the \citetalias{conroy_early-type_2014} comparison sample. Thus, we use results from the stacked SDSS spectra instead of hierarchical Bayes. The problems introduced by stacking are often worse with low numbers of contributing spectra (e.g. LEGA-C), and therefore the SDSS stacks, each with thousands of spectra, are more robust to stacking. Following the discussion in Section~\ref{sec:hbm}, hierarchical Bayesian analysis is preferred over stacking in almost all cases unless the cost of fitting becomes prohibitively expensive. Additionally, for extremely noisy data where fits to individual galaxies are impossible or untrustworthy, stacking may be preferred.

In Figure~\ref{fig:ind_res} we show the results of the hierarchical Bayesian modeling (blue) in comparison to the results from the individual fits (gray). This figure serves as an illustration and shows that our method successfully captures the average abundances and ages in each $\sigma_v$ bin. The actual trends and their significance (for these and other elements) are discussed in the next section. In Appendix~\ref{app:sim_stack} we compare hierarchical Bayesian modeling to the stacking method used in \citetalias{conroy_early-type_2014} and find mostly consistent results, although the Bayesian method is more accurate and the uncertainties are better estimated. The results of these tests are summarized in Figure~\ref{fig:sim_stack}. In Appendix~\ref{app:HBM} we provide a more complete derivation of the hierarchical Bayesian method and its implementation.

\section{Abundance Patterns of \texorpdfstring{$\lowercase{z}\sim0.7$}{z0.7} Galaxies}
\label{sec:results}

\begin{figure*}
    \centering
    \includegraphics[width=0.9\textwidth]{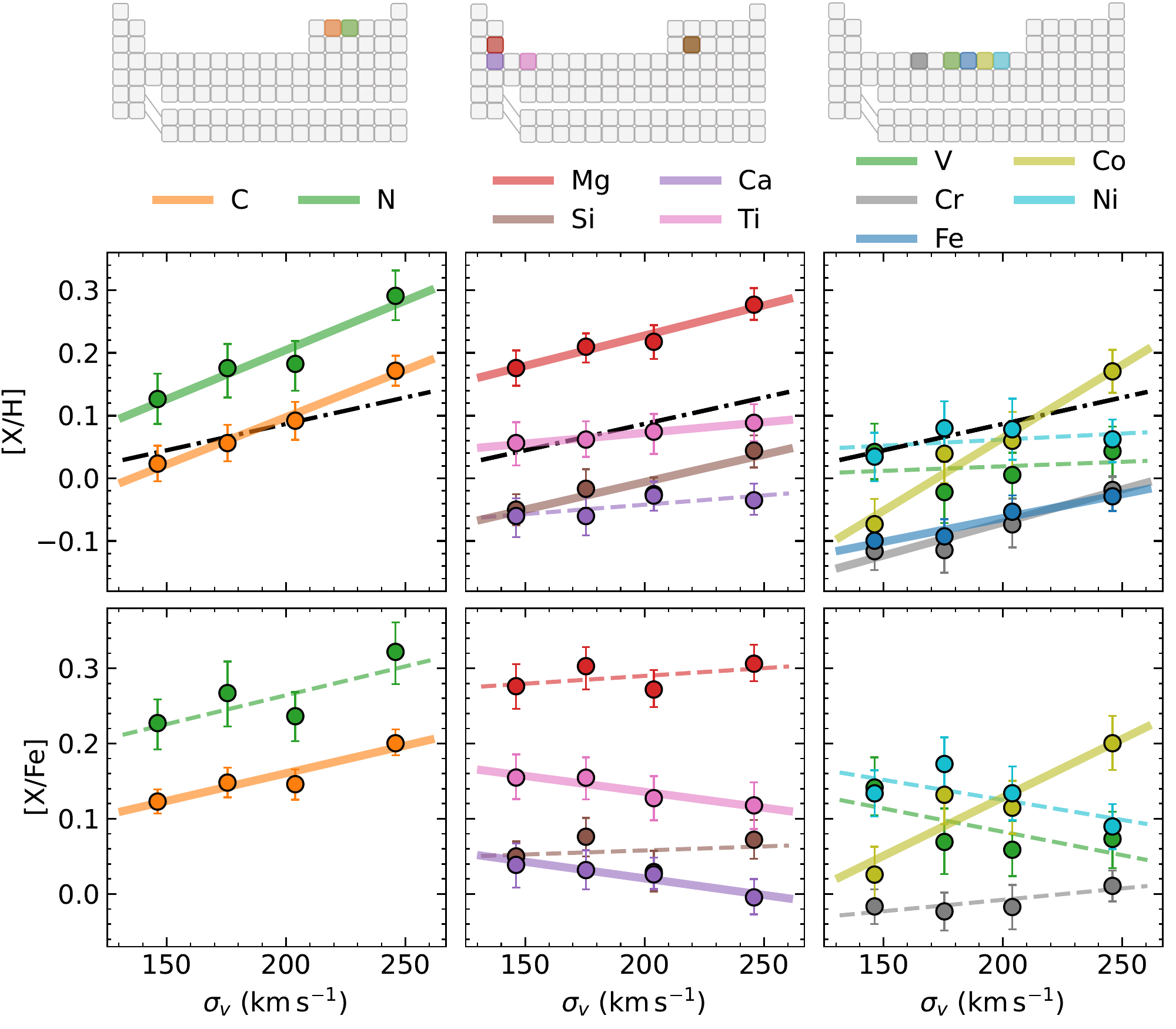}
    \caption{Absolute abundances [X/H] (top row) and abundance ratios [X/Fe] (bottom row) of C, N (left column), $\alpha$ elements (middle column), and iron peak elements (right column) as a function of observed velocity dispersion. The galaxy sample was split into four velocity dispersion bins before fitting. The individual points represent the mean abundance of the population of galaxies in each bin $(\mu_{\theta_i})$, as defined in Section \ref{sec:hbm}. The error bars reflect the $1\sigma$ 16th and 84th percentiles of the $(\mu_{\theta_i})$ PDFs. For each element, we include the best-fit linear relation to the bins. Solid lines indicate elements with slopes that deviate significantly from zero (at the 3$\sigma$ level), whereas elements with dashed lines do not. The dash-dotted black lines in the [X/H] panels reflect the ``total'' metallicity [Z/H], which we define as the sum of all ten absolute abundances included in the fits.}
    \label{fig:ab_vs}
\end{figure*}

\begin{figure}
    \centering
    \includegraphics[width=0.95\columnwidth]{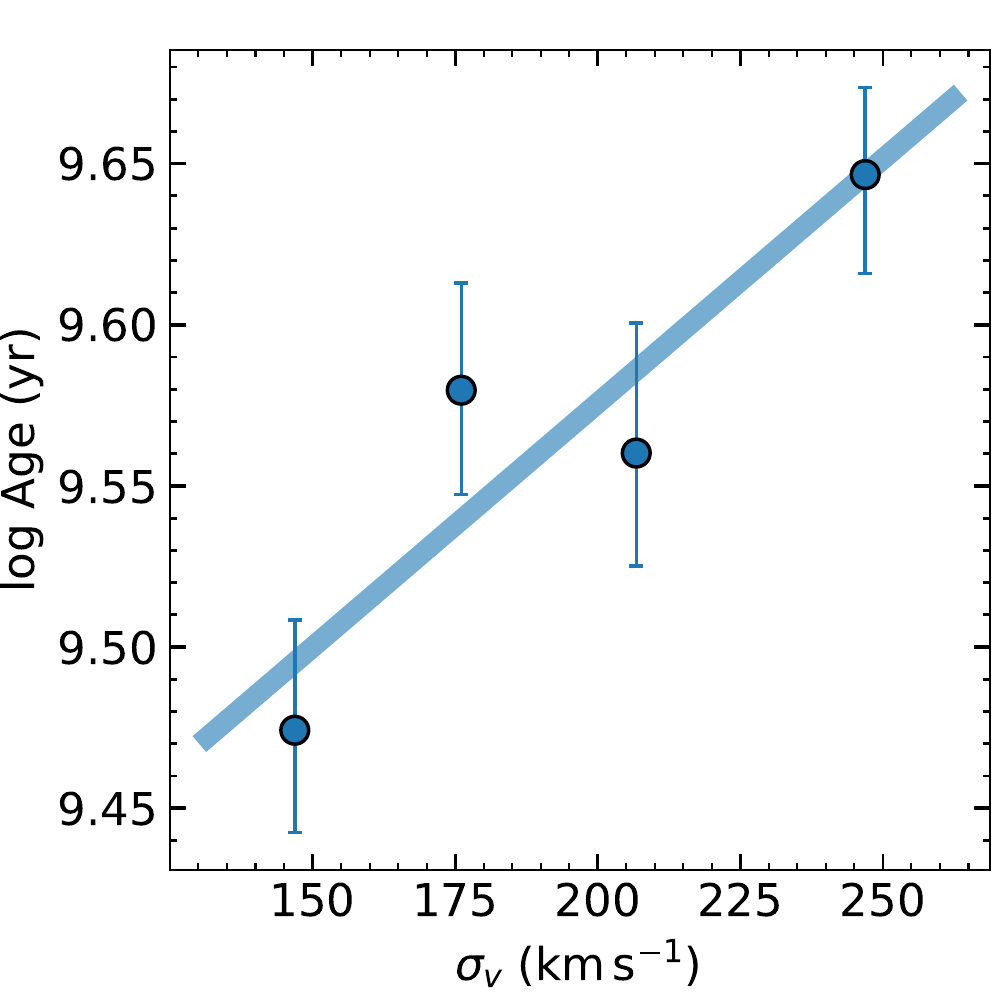}
    \caption{Stellar population ages as a function of observed velocity dispersion. The best-fit line from least-squares fitting is shown. It is clear that galaxies with higher velocity dispersions formed earlier.}
    \label{fig:age_v_sig_simp}
\end{figure}

\begin{figure*}
    \centering
    \includegraphics[width=\textwidth]{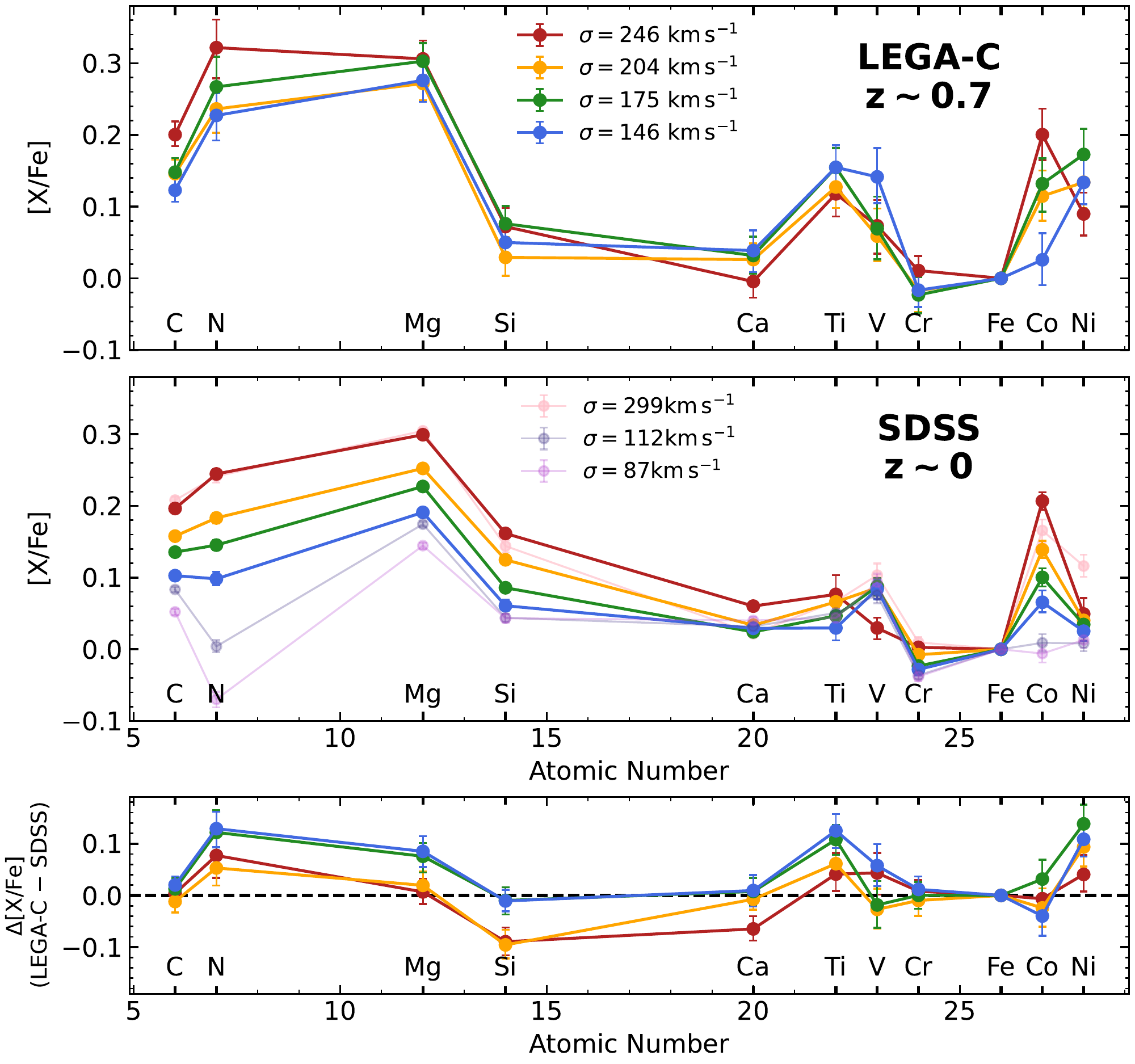}
    \caption{The elemental abundance patterns derived from the LEGA-C $z\sim0.7$ sample (top panel) and the SDSS $z\sim0$ sample from \citetalias{conroy_early-type_2014} (middle panel). The SDSS sample covers velocity dispersions ranging from $\sigma=90$ $\rm{km\;s^{-1}}$ (light purple) to $\sigma=300$ $\rm{km\;s^{-1}}$ (pink). We highlight the best matching dispersion bins in LEGA-C and SDSS using lines of the same color. The abundance patterns are remarkably similar. In the bottom panel we show the difference in abundance ratios between LEGA-C and SDSS. It is apparent that the N, Mg, Ti, and Ni abundance ratios are enhanced in LEGA-C compared to SDSS, especially at low dispersions.}
    \label{fig:abpattern}
\end{figure*}

The elemental abundance results as a function of observed velocity dispersion based on the hierarchical Bayesian modeling are summarized in Figure~\ref{fig:ab_vs}. The ten elements are split into three categories: CNO products (C, N; left column), $\alpha$ elements (Mg, Ca, Si, Ti; middle column), and Fe-peak elements (V, Cr, Co, Fe, Ni; right column). In the top row we show the absolute abundances [X/H] as a function of observed velocity dispersion, while in the bottom row we show the abundance ratios [X/Fe]. We evaluate the correlation with least-squares minimization and show the best-fit line for each element. The dashed black line in each panel of the top row reflects the best-fit line to the ``total'' metallicity [Z/H] which is defined as the sum of all ten absolute abundances. 

The absolute abundances of each element, as well as the total metallicity [Z/H], scale positively with velocity dispersion. All elements, except Ca, V, and Ni, have positive slopes that are significant at the $3\sigma$-level. The trend between velocity dispersion and absolute abundance has been studied extensively at low-$z$ \citep[e.g.,][]{gallazzi_ages_2005, thomas_environment_2010, conroy_early-type_2014}, this is the first time we show this trend at intermediate redshift for as many elements. 

The positive trend between the absolute abundances and velocity dispersion is in broad agreement with what is found at $z\sim0$ \citep[e.g.][]{gallazzi_ages_2005, thomas_environment_2010, conroy_early-type_2014, mcdermid_atlas3d_2015}. It is also consistent with initial work at $z\sim0.5$ that show results for [Z/H], [Fe/H], [Mg/H], [C/H], [N/H], and [Ca/H] \citep{gallazzi_charting_2014, choi_assembly_2014, leethochawalit_evolution_2019}. These trends have historically been interpreted as galaxies with larger velocity dispersions having steeper gravitational potential wells, thus allowing them to hold onto metal-rich gas during feedback events like supernovae and high-speed stellar winds \citep{larson_effects_1974, dekel_origin_1986, tremonti_origin_2004}. In Section \ref{sec:discussion} we discuss whether observed velocity dispersion is indeed a good tracer of the gravitational potential. 

We now turn to the Fe-scaled abundance ratios [X/Fe] as a function of velocity dispersion. The abundance ratio [$\alpha$/Fe] is often used as a tracer of the star-formation timescale, as $\alpha$ elements are produced on shorter timescales than Fe due to the relative difference in timescales between core-collapse and Type Ia SNe \citep[e.g.][]{matteucci_abundance_1994, trager_stellar_2000, thomas_epochs_2005,conroy_early-type_2014}. A similar argument can be used for [C/Fe] and [N/Fe], with C and N probing a slightly longer timescale than $\alpha$-elements due to the contributions from AGB winds. In the second row of Figure~\ref{fig:ab_vs} we find that only [Co/Fe] and [C/Fe] have significant ($3\sigma$-level) positive correlations with $\sigma_v$, while the $\alpha$ elements Ti and Ca have significant mild negative correlations. All of the other elements have abundance ratios consistent with no correlation. To first order, the mostly flat [X/Fe] results imply that the average star-formation timescales are similar and we see no significant trend with velocity dispersions over the studied range. In other words, the gravitational potential of a galaxy appears to regulate absolute abundances but \textit{not} the star-formation timescale.

The above abundance ratio results are mostly consistent with \citet{choi_assembly_2014} -- the only other study of abundance patterns at $z>0$. They presented the elemental abundances C, N, Mg, Fe, and Ca for stacks of massive quiescent galaxies in six different redshift bins out to $z\sim0.7$, though their $0.55<z<0.7$ stack has S/N much less than what is presented here ($25-150\,$\AA for \citet{choi_assembly_2014} compared to $\approx60$\,\AA for each individual LEGA-C spectrum). They find mostly flat elemental abundance ratios with stellar mass, especially in the higher redshift bins. The flat [X/Fe] results presented here and in \citet{choi_assembly_2014} are somewhat surprising, as at $z\sim0$, massive quiescent galaxies appear to have a positive correlation between abundance ratios and stellar mass or velocity dispersion \citep[e.g.][]{worthey_individual_2014, conroy_early-type_2014, mcdermid_atlas3d_2015}. However, the limited dynamic range in velocity dispersion and relatively small sample size make it difficult to say for certain whether the above abundance trends with velocity dispersion exist at this redshift. Thus a larger statistical sample that reaches to lower stellar masses and velocity dispersions would be necessary to directly test this correlation past $z\sim0$.

The bottom row of Figure~\ref{fig:ab_vs} also gives some insights into the nucleosynthetic origins of the elements. While N does not have a significantly positive slope due to the large uncertainties on the binned measurements, it still appears to trace [C/Fe] as a function of velocity dispersion. The similarity in slope may indicate that they are forged from the same processes and on similar timescales. This is in qualitative agreement with results at $z\sim0$ \citep{graves_ages_2007, schiavon_population_2007, smith_abundance_2009, johansson_chemical_2012}. The slightly negative slopes found for [Ti/Fe] and [Ca/Fe] may indicate Ti and Ca are not pure core-collapse products and could have significant contributions from both core-collapse and Type Ia supernovae. In the local universe, this is also found to be the case \citep{saglia_puzzlingly_2002, cenarro_near-infrared_2003, thomas_stellar_2003, graves_ages_2007, schiavon_population_2007, smith_abundance_2009, johansson_chemical_2012}. Finally, the Fe-peak element Co exhibits a steep positive correlation with velocity dispersion. This result is surprising, as no other element has such an extreme trend with velocity dispersion. Interestingly, the strong trend with Co is also found in massive early-type galaxies in the nearby universe \citep{conroy_early-type_2014}. As explained by \citet{conroy_early-type_2014}, this result could indicate that a significant portion of Co may originate from core-collapse supernovae, and not Type Ia. Indeed, this is seen in Type Ia SNe yields of \citet{nomoto_accreting_1984}, who show a deficit of Co compared to other Fe-peak elements.

It is not obvious why only [C/Fe] and [Co/Fe] show significant positive relations with $\sigma$, while none of the $\alpha$ elements do. If the star-formation timescale is the primary driving factor setting the slopes of the abundance trends, then core-collapse products have the steepest trend with velocity dispersion, followed by CNO and Type Ia products. The fact that we do not observe these trends, and instead find Co (an Fe-peak element) to have the steepest slope, followed by C (AGB and core-collapse product) may indicate there are other factors at play. One such explanation could be that the IMF varies with velocity dispersion \citep[as found by e.g.][]{van_dokkum_substantial_2010, conroy_slar_2012, conroy_dynamical_2013}. Metallicity- and mass-dependent supernova yields combined with a more top-heavy IMF can alter the relative abundances of the elements. Further chemical evolution modeling is required to confirm whether a variable IMF could explain the observed abundance trends. We caution that the observed trends are measured to varying degrees of significance and larger samples are required to confirm all trends.
\begin{figure*}
    \centering
    \includegraphics[width=\textwidth]{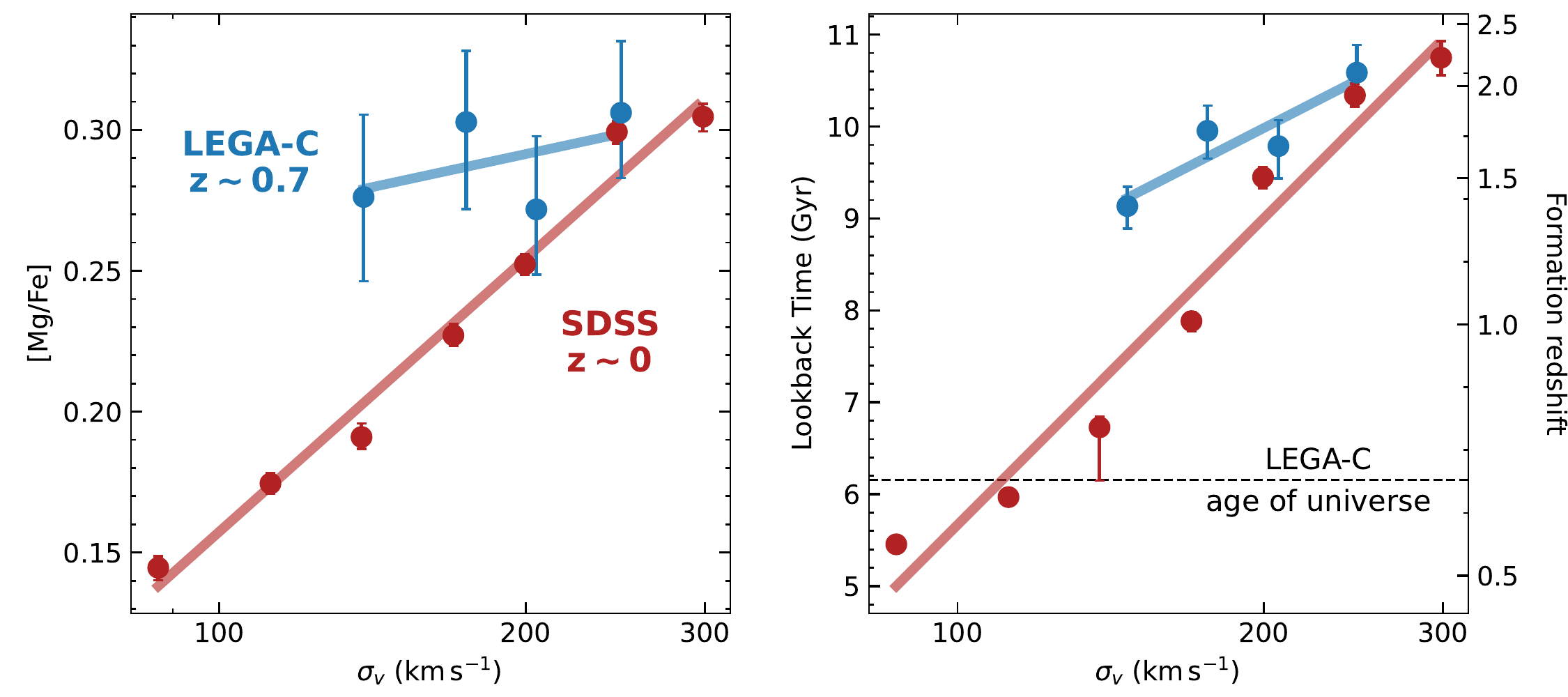}
    \caption{Comparison of the [Mg/Fe] \textit{(left)} and formation redshift \textit{(right)} for stacks of massive quiescent galaxies in SDSS at $z\sim0$ \citepalias[red;][]{conroy_early-type_2014} and LEGA-C at $z\sim0.7$ (blue), along with the best-fit power laws. The formation redshift is calculated by taking into account the age of the universe at each redshift. In both samples, galaxies with higher velocity dispersions have higher [Mg/Fe] and formed earlier. The $z\sim0.7$ and $z\sim0$ populations differ most significantly in the lowest velocity dispersion bin, with the $z\sim0.7$ galaxies having $\approx0.1\,\rm{dex}$ higher [Mg/Fe] and $\approx2\,$Gyr earlier formation times compared to the corresponding $z\sim0$ bin. This difference can be explained by newly quenched galaxies with young ages and lower [Mg/Fe] being added to the low-$\sigma_v$ bins over the last six billion years.}
    \label{fig:age_v_sig}
\end{figure*}

Next we turn to the SSP-equivalent ages as a function of velocity dispersion. In Figure~\ref{fig:age_v_sig_simp} we find a significant positive trend between age and velocity dispersion. This result implies that galaxies with larger gravitational potentials form the bulk of their stars at earlier epochs. The trend between stellar age and velocity dispersion is in agreement with other studies at similar redshift that look at age as a function of stellar mass \citep[e.g.][]{gallazzi_charting_2014,choi_assembly_2014,leethochawalit_evolution_2019,beverage_elemental_2021} and with studies of massive quiescent galaxies at $z\sim0$ \citep{thomas_environment_2010, mcdermid_atlas3d_2015, barone_sami_2018}.

\section{Discussion}
\label{sec:discussion}
\subsection{Comparison with local early-type galaxies}

There are a few key advantages to studying the chemical compositions of galaxies at higher redshift. First, their in-situ stellar populations experience less pollution by late-time mergers. Second, observations at higher redshift reach closer to the epoch of formation, making the ages and star-formation histories easier to determine. And third, by comparing these results to $z\sim0$ we can further characterize the evolution of the quiescent galaxy population over cosmic time.

In this section we first compare our $z\sim0.7$ abundance results to stacks of massive local early-type galaxies from SDSS and to test whether the massive quiescent galaxy population has evolved over the past six billion years. Figure~\ref{fig:abpattern} shows the abundance patterns for both the LEGA-C (top panel) and SDSS (middle panel) stacks as a function of velocity dispersion, along with the residuals between the two populations (bottom panel). As a reminder, the SDSS stacks were taken directly from \citetalias{conroy_early-type_2014} but re-fit here using the same SSP models and \texttt{alf} setup as the individual LEGA-C spectra (see section~\ref{sec:alf}). The SDSS stacks cover a larger range in velocity dispersion (90\;km\,s$^{-1}$ to 300\;km\,s$^{-1}$) compared to the LEGA-C sample (150\;km\,s$^{-1}$ to 250\;km\,s$^{-1}$) and thus we highlight the best matching dispersion bins in LEGA-C and SDSS using lines of the same color.

One concern in comparing the SDSS and LEGA-C results is that the SDSS spectra have an aperture that covers the inner $\sim0.5\;$R$_{e}$, while the LEGA-C aperture covers $\sim1\;$R$_{e}$. Thus, the presence of stellar population gradients in these galaxies can significantly bias the resulting measurements integrated over the aperture. Spatially resolved studies of massive early-type galaxies in the local universe reveal steep [X/H] gradients but flat stellar age and abundance ratio [X/Fe] gradients \citep[e.g.][]{spolaor_early-type_2010, greene_massive_2019, feldmeier-krause_stellar_2021, parikh_sdss-iv_2021}. \citet{choi_assembly_2014} model the effects of metallicity gradients on integrated [Fe/H] measurements and find only a modest effect (a difference of only 0.05$\,$dex between fibers covering $0.5\;$R$_{e}$ and $1\;$R$_{e}$). Nonetheless, we refrain from comparing absolute abundances [X/H], and instead focus on abundance ratios [X/Fe] and ages, which are less impacted by aperture bias.

Figure~\ref{fig:abpattern} shows that the SDSS and LEGA-C abundance patterns closely trace one another. In both cases CNO products (C, N) and $\alpha$-element Mg all have enhanced abundances followed by a steep drop to approximately solar abundances for the heavier $\alpha$ (Si, Ca, Ti) and Fe-peak (V, Cr, Fe, Co, Ni) elements. These similarities imply that on average, massive quiescent galaxy at $z\sim0.7$ have similar chemical enrichment histories to those at $z\sim0$. One difference in the $z\sim0$ and $z\sim0.7$ abundance patterns is that Si is slightly depressed in LEGA-C compared to SDSS, while Ti is slightly enhanced. The small uncertainties ($\lesssim0.05$\,dex for both SDSS and LEGA-C) on these measurements suggest that these abundance differences, albeit subtle, are real. Chemical evolution modeling is required to shed light on the origin of these differences.

Another difference between the SDSS and LEGA-C abundance patterns is that the lowest velocity dispersion bins in SDSS have depressed abundance ratios compared to LEGA-C, in particular for N, Mg, Ti, and Ni (see bottom panel of Figure~\ref{fig:abpattern}). As such, the trends between [X/Fe] and velocity dispersion for these elements become stronger over the past six billion years. In the left panel of Figure~\ref{fig:age_v_sig} we show this directly by comparing [Mg/Fe] for the $z\sim0$ and $z\sim0.7$ populations. It is clear that the $z\sim0.7$ population is Mg-enhanced compared to the quiescent galaxies at $z\sim0$, especially at the lowest velocity dispersions. We discuss possible implications of this result in the next section. 

Finally, in the right panel of Figure~\ref{fig:age_v_sig}, we compare the formation redshift of the $z\sim0$ and $z\sim0.7$ quiescent galaxy populations. We calculate the formation redshift by correcting the stellar ages for the age of the universe at each redshift. Thus, the right panel Figure~\ref{fig:age_v_sig} allows us to directly compare the stellar ages of the galaxies in the two redshift bins. In both cases we find that galaxies with the largest velocity dispersions on average formed the earliest. We also find that the $z\sim0.7$ quiescent galaxy population on average formed earlier than $z\sim0$ quiescent galaxies, especially at lower dispersions (by $\approx2$\;Gyr for $\sigma=150\;\rm{km\,s^{-1}}$). Thus, both the stellar ages \textit{and} the abundance ratios differ the most in the lowest velocity dispersion bins. This difference in the lowest dispersion bin could be explained by mergers or newly quenched galaxies being added to the quiescent population at later times. Alternatively, some star formation between $z\sim0.7$ and $z\sim0$ may explain the formation time differences. In particular, because SSP-equivalent ages are basically luminosity-weighted ages, and thus they are more sensitive to the younger stellar populations.

There are several additional caveats that must be considered before interpreting the abundance and age results in the context of galaxy evolution. First, we assume that the observed velocity dispersion traces the strength of the potential well. However, \citet{bezanson_spatially_2018} find many of the quiescent galaxies in LEGA-C are in fact supported by rotation. As a result, the measured velocity dispersions may underestimate the strength of the potential well if the disks have significant rotational components ($V/\sigma\sim1$) and are also observed face-on. Therefore, the observed trends with velocity dispersion could be artificially flattened compared to SDSS, where rotation appears to be less prevalent. We investigate the significance of this effect using the inclination- and aperture-corrected velocity dispersions provided by \citet[][Eq. 2]{van_der_wel_large_2021}. The abundance ratio trends remain qualitatively the same and thus we conclude that inclined rotating disks have minimal impact on our results.

A second caveat is that the SDSS selection is different than our LEGA-C selection; \citetalias{conroy_early-type_2014} remove galaxies that fall outside of the central slice of the fundamental plane. Effectively, this selection removes inclined rotation-supported disks. We test the impact of this effect by selecting a sub-sample of LEGA-C galaxies with axis ratios $b/a>0.6$, which removes the inclined disks from the sample. We find that the trends with abundance ratio remain qualitatively the same. Therefore, this selection criterion appears to have little impact on our results.
Finally, SDSS galaxies are selected to be quiescent based on H$\alpha$ and [OII]$\lambda3727$, whereas LEGA-C quiescent galaxies are selected by their $UVJ$ colors. As mentioned earlier (Section~\ref{sec:sdss_compare}), most LEGA-C galaxies still have some [OII] emission, which most likely originates from blue evolved stars, and thus may decline as the stellar population ages to present day. Hence, we do not expect that the difference in selection criteria has any significant affect on the comparison.

\subsection{Implications for galaxy evolution since \texorpdfstring{$\lowercase{z}\sim0.7$}{z0.7}}

In the previous section we found that massive quiescent galaxies at $z\sim0.7$ stellar have ages and abundance ratios broadly consistent with observations at $z\sim0$. However, galaxies in the lowest velocity dispersion bin are found to have formed earlier and have more enhanced abundance ratios compared with galaxies at $z\sim0$. This result agrees with the findings of \citet{choi_assembly_2014} who find abundance ratios and stellar ages that are consistent with passive evolution since $z\sim0.7$ for the \textit{most} massive ($\log\rm{M}>11\,\rm{M_\odot}$ or $\sigma_v\approx200\,\rm{km\,s^{-1}}$) quiescent galaxy population. In this section we discuss possible evolutionary scenarios that could explain the results at low velocity dispersion. 

Quiescent galaxy populations (that by definition no longer form stars) can still evolve via minor mergers \citep[e.g.][]{bezanson_relation_2009,naab_minor_2009,hopkins_compact_2009, van_de_sande_stellar_2013}, the addition of newly quenched galaxies to the quiescent population \citep[e.g.,][]{van_dokkum_morphological_2001, carollo_newly_2013, poggianti_evolution_2013, belli_mosfire_2019}, and late-time star formation \citep[e.g.][]{donas_galex_2007,schawinski_effect_2007, thomas_environment_2010}. In the minor merger scenario, the abundance ratios and ages of massive quiescent galaxies are diluted by the accretion of lower mass galaxies (with mass ratios of around 1:10). Due to the younger ages and near-solar abundances of lower mass galaxies, minor mergers typically lower the abundance ratios of massive quiescent galaxies over time. Furthermore, they result in radial stellar population gradients, as have been found in massive elliptical galaxies in the local universe \citep[e.g.,][]{greene_massive_2015, greene_massive_2019, oyarzun_signatures_2019, feldmeier-krause_stellar_2021, parikh_sdss-iv_2021}, and perhaps even at earlier times \citep{suess_color_2020}. However, due to prevalence of galactic conformity, or the idea that old galaxies merge with other old galaxies \citep[e.g.,][]{weinmann_properties_2006, kauffmann_re-examination_2013}, merging would not significantly impact stellar ages. 

The addition of newly quenched galaxies can also reduce the average ages and abundance ratios of quiescent galaxies over time, as is found in Figure~\ref{fig:age_v_sig}. In this scenario, star-forming galaxies continue to quench after $z\sim0.7$, especially at lower stellar masses. Thus, the newly added young and Fe-enhanced galaxies lower the \textit{average} ages and abundance ratios. We do indeed know that the quiescent galaxy population has grown significantly since $z\sim1$ \citep[e.g.,][]{bell_nearly_2004, faber_galaxy_2007, taylor_masses_2010, muzzin_evolution_2013,tomczak_galaxy_2014,mcleod_evolution_2021}. Finally, late-time star-formation, also known as ``rejuvenation,'' can alter the abundances and ages of the quiescent galaxy population by giving more weight to the newly formed bright, young, and more chemically evolved stellar population.

All of the above evolutionary scenarios produce the observed decrease in formation redshift and abundance ratios over time found in Figure~\ref{fig:age_v_sig}. However, in order to explain this decrease, the evolution must be the strongest in the lowest velocity dispersion bin. Both progenitor bias and late-time star-formation -- but not minor mergers -- preferentially affect the evolution of galaxies in the lower velocity dispersion bins. For progenitor bias, this is because the most massive galaxies are already quiescent by $z\sim0.7$ \citep{mcleod_evolution_2021, taylor_velocity_2022}. Thus, evolution of the quiescent galaxy population via quenching of star-forming galaxies between $z\sim0.7$ and $z\sim0$ is strongest for galaxies with lower velocity dispersions. Alternatively, evolution through late-time star-formation could explain the results if this scenario preferentially affects galaxies with lower velocity dispersions. Indeed, in the local universe, \citet{thomas_environment_2010} find that star formation``rejuvenation'' is more commonly observed in galaxies with lower stellar masses. However, studies of the star-formation histories of LEGA-C galaxies at $z\sim0.7$ find that rejuvenation is not necessarily an important evolutionary channel for the growth of the quiescent population since $z\sim0.7$ \citep{chauke_rejuvenation_2019}.

One way to differentiate between the progenitor bias and rejuvenation scenarios would be to trace the scatter of individual abundances and ages over time; progenitor bias adds young galaxies to the existing old population and thus increases the scatter in individual galaxies over time, whereas late-time star formation does not. With deeper spectra that trace to lower velocity dispersions, as well as careful treatment of S/N-related scatter, such analysis will be possible.

\subsection{Comparison with theoretical predictions}

\begin{figure}
    \centering
    \includegraphics[width=\columnwidth]{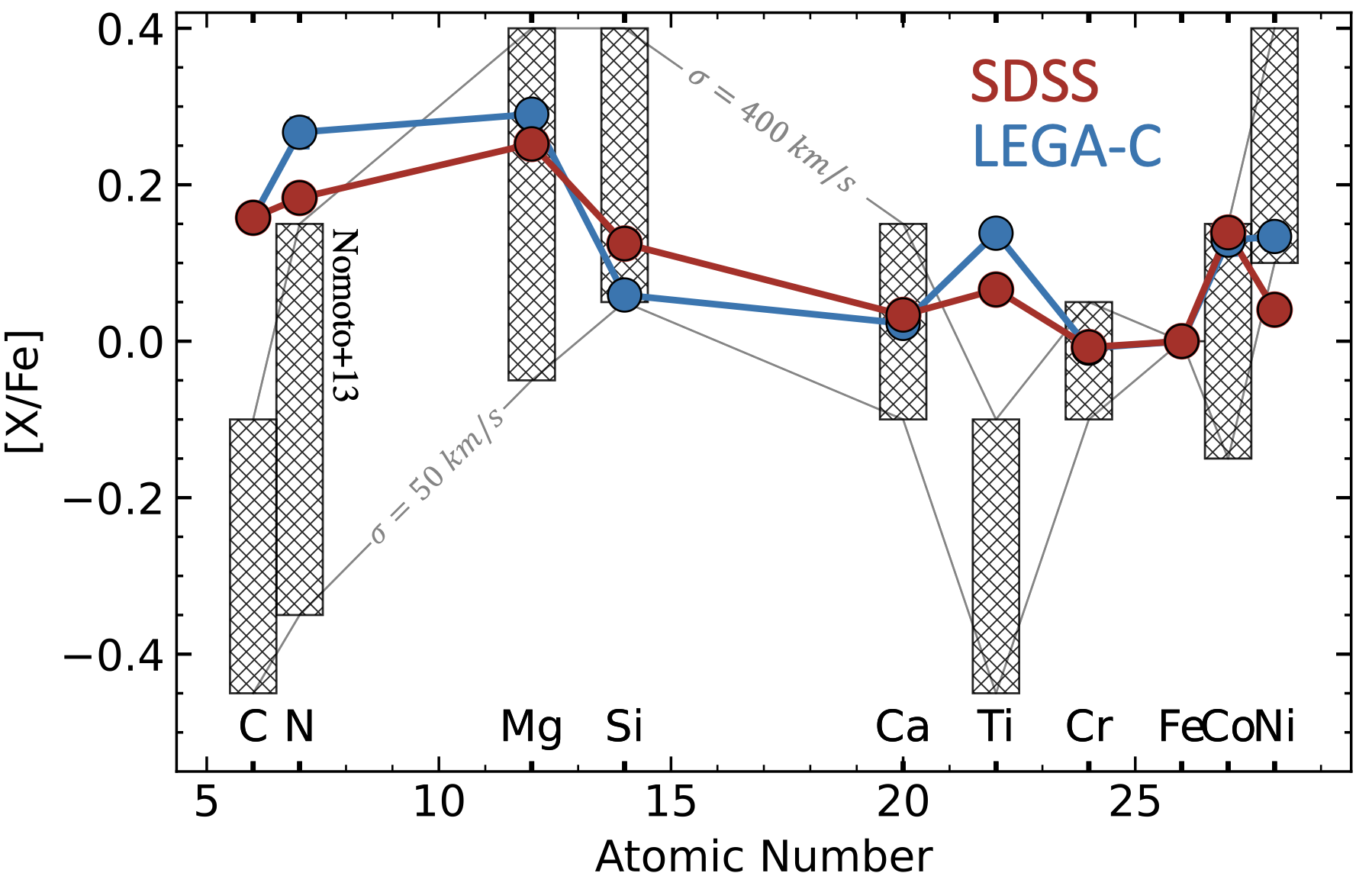}
    \caption{Comparison between the observed abundance patterns and theoretical predictions from one-zone chemical evolution models for elliptical galaxies from \citet{nomoto_nucleosynthesis_2013}. The blue line represents the average abundance pattern of the entire LEGA-C sample (with an average velocity dispersion of $\sigma=200$~\kms) and the red line represents the SDSS abundance pattern for a bin centered around $\sigma=200$~\kms. The gray lines show the predicted abundance patterns for elliptical galaxies with $\sigma=50$~\kms\, and $\sigma=400$~\kms\, from chemical evolution modeling. We shade the region between these lines for each element to show the predicted range of abundance ratios. \citet{nomoto_nucleosynthesis_2013} assume an infall+wind model with a Salpeter IMF, and a star formation history that reflects the ``red and dead'' properties of local elliptical galaxies. There are major differences between the observations and the theoretical predictions, especially for elements C, N, and Ti.}
    \label{fig:nomoto}
\end{figure}

The elemental abundance patterns provide unique constraints on chemical enrichment histories of galaxies. In this section we compare our results with theoretical predictions from \citet{nomoto_nucleosynthesis_2013}. They report theoretical predictions for 16 elements in [X/Fe]-[Fe/H] for two elliptical galaxy models -- one low-mass galaxy with $\sigma=50$~\kms, and one high-mass galaxy with $\sigma=400$~\kms -- from a one-zone chemical evolution model. This model assumes infall of primordial gas, supernova winds, a \citet{salpeter_luminosity_1955} IMF, and star formation histories that reflect the observed ages and formation timescales of elliptical galaxies in the local universe. In Figure~\ref{fig:nomoto} we show the results of these models at a fixed Fe-abundance ([Fe/H]$\sim0$) (gray lines), and shade the regions between the two abundances. For comparison, in blue we show the average abundance pattern for \textit{all} galaxies in the $z\sim0.7$ quiescent galaxy, along with the SDSS abundance pattern at $\sigma\sim200$~\kms (to match the average velocity dispersion of the $z\sim0.7$ sample) in red. [V/Fe] is excluded from this figure, as this element is not included in the analysis presented by \citet{nomoto_nucleosynthesis_2013}.


This figure shows that the observed abundance patterns are very different from those predicted by chemical evolution models. In particular, the models cannot recover C, N, and Ti. C and N are both elements with large contribution from evolved intermediate-mass stars and thus the models may require longer star-formation histories to incorporate more CNO products, or alternatively require different supernovae and AGB yields. Interestingly, chemical evolution studies in the Local Universe find that AGB production of N has a shorter time-delay than previously thought, and could explain some of the observed discrepancy \citep{johnson_empirical_2022}. Furthermore, the under-abundance of Ti predicted by models has been noted in the Solar Neighborhood \citep{nomoto_nucleosynthesis_2006}, the Milky Way halo \citep{francois_evolution_2004}, and dwarf galaxies \citep{kirby_multi-element_2011}, and is likely due to uncertain core-collapse yields. The deviation of our observations from the theoretical predictions by this model imply that there may be components of the chemical enrichment of massive quiescent galaxies that are still poorly understood. However, here we make the comparison with only one chemical evolution model. In the future with a wider set of models and higher-S/N data, this sort of comparison may be able to provide more concrete constraints on the IMF and yields of massive quiescent galaxies beyond the low-redshift universe.


\section{Conclusion}
\label{sec:conclusion}

In this paper we presented detailed elemental abundance patterns and stellar population ages for 135 massive quiescent galaxies at $z\sim0.7$ drawn from the LEGA-C survey. We modeled the ultra-deep rest-frame optical spectra using a full-spectrum fitting code with varying elemental abundances. We then placed the galaxies in four bins of velocity dispersion and calculated the average abundance results in each bin using a hierarchical Bayesian modeling technique. Finally, we compared the binned results to stacks of massive early-type galaxies at $z\sim0$. The main conclusions are summarized as follows:

\begin{enumerate}
    \item The absolute abundances [X/H] are correlated with the observed velocity dispersion, such that galaxies with larger $\sigma_v$ are more metal-rich. This result reinforces that galaxies with larger gravitational potentials are better at retaining their enriched gas reservoir. 
    \item The abundance ratios [X/Fe] show only mild or non-significant trend with velocity dispersions for nearly all elements. To first order this result implies that, on average, massive quiescent galaxies form on similar timescales over the studied range of velocity dispersions. Briefly, we note the small range in velocity dispersions probed by this study. A larger sample probing lower-dispersion galaxies may reveal a possible trend.
    \item The stellar population age increase as a function of velocity dispersion such that galaxies with deeper gravitational potentials formed earlier.
    \item To first order, massive quiescent galaxies at $z\sim0.7$ and $z\sim0$ have remarkably similar abundance patterns, with enhanced CNO products (C, N) and $\alpha$-element Mg followed by a steep drop to approximately solar abundances for the heavier $\alpha$ (Si, Ca, Ti) and Fe-peak (V, Cr, Fe, Co, Ni) elements. However, in the lower velocity dispersion bins, galaxies at $z\sim0.7$ are found to have formed earlier and are more enhanced in [X/Fe] for elements N, Mg, Ti, and Ni compared to $z\sim0$. These findings may indicate that the low-dispersion quiescent galaxy population is still evolving either by late-time star formation or the late additions of newly quenched galaxies between $z\sim0.7$ and $z\sim0$.
    \item The measured abundance patterns show major differences with theoretical predictions based on one-zone chemical evolution models for elliptical galaxies from \citet{nomoto_nucleosynthesis_2013}, indicating that our current understanding of the detailed chemical enrichment histories of massive quiescent galaxies is still limited.
\end{enumerate}

The LEGA-C survey has enabled us to measure the detailed and robust elemental abundance patterns of quiescent galaxies beyond $z\sim0.5$. These abundance patterns may serve as powerful probes of the chemical enrichment and formation histories of these galaxies. However, the interpretation is currently still limited by low S/N measurements, our reliance on ``stacking'' spectra, and our incomplete theoretical understanding of the nucleosynthetic origins of many elements. Using more detailed and updated chemical evolution modeling, along with abundance pattern measurements for larger samples extending to higher redshift enabled by JWST, this field is expected to make major leaps forward over the coming decade.\\

We thank Adam Carnall, Anna de Graaff, Dan Foreman-Mackey, Jenny Greene, Meng Gu, Benjamin Johnson, Chiaki Kobayashi, and David Weinberg for useful discussions. We also thank the LEGA-C team for making their dataset public. We acknowledge support NSF AAG grants AST-1908748 and 1909942. A.G.B. is supported by the National Science Foundation Graduate Research Fellowship Program under grant No. DGE 1752814 and DGE 2146752.

\software{\texttt{GALFIT} \citep{peng_detailed_2010}, FSPS \citep{conroy_propagation_2009}, \texttt{FAST} \citep{kriek_ultra-deep_2009}, \texttt{alf} \citep{conroy_slar_2012,conroy_metal-rich_2018}, \texttt{PPXF} \citep{cappellari_parametric_2004,cappellari_full_2022}}

\bibliography{zoterolib}
\appendix
\section{Parameter Retrieval From Simulated Spectra}
\label{app:simulation}

The abundance results in this paper are evaluated using a hierarchical Bayesian model. The first level of this model relies on the individual fits to the LEGA-C spectra. The posteriors of these fits are then combined in a Bayesian framework as described in Section~\ref{sec:hbm}. Thus, for reliable combined results we require the individual parameters to be accurate and constrained to at least within the imposed priors. In this appendix we describe the simulation and recovery test used to quantitatively evaluate which parameters can be constrained.

\begin{figure*}[b!]
    \centering
    \includegraphics[width=\textwidth]{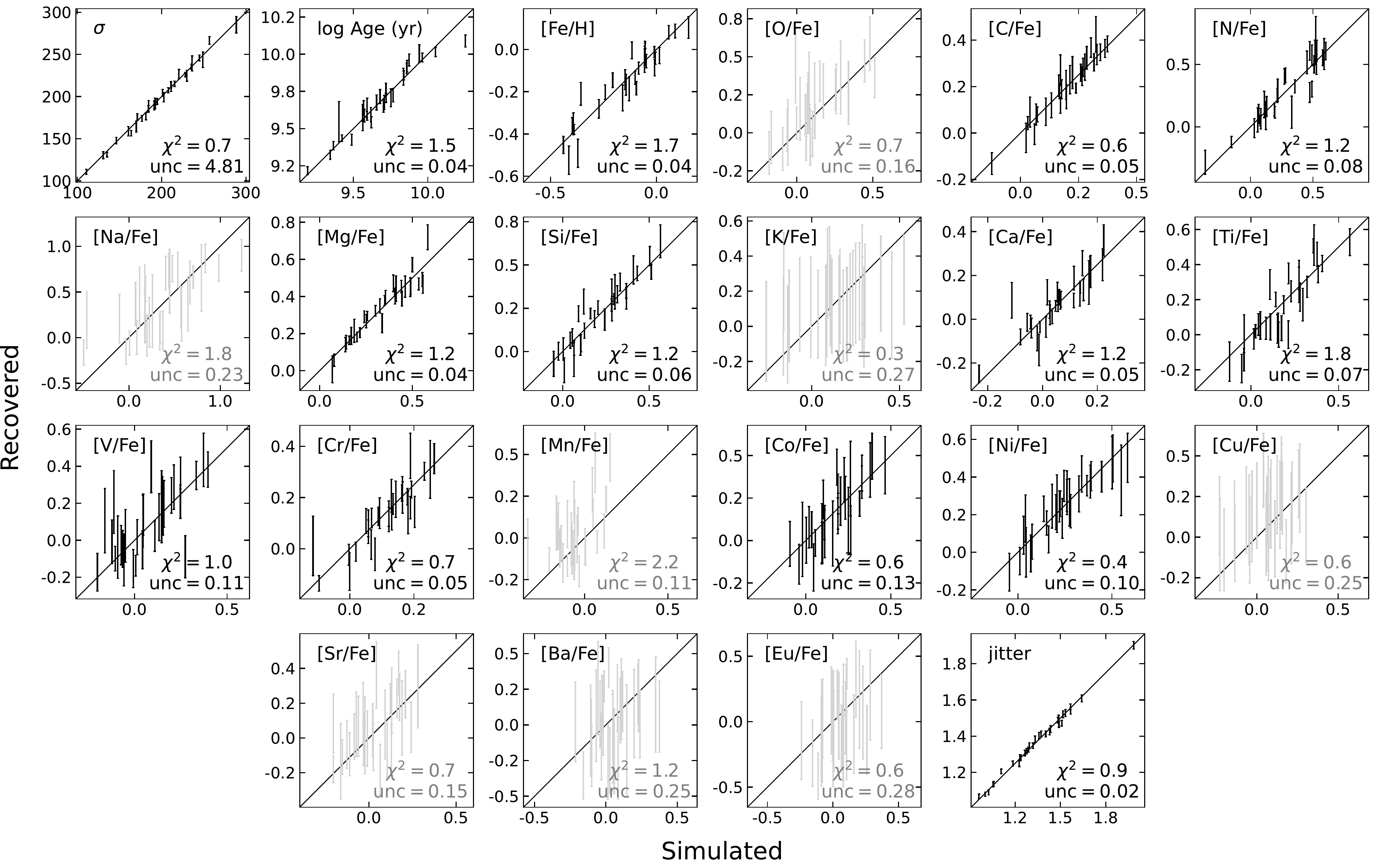}
    \caption{Recovery results from simulating 30 LEGA-C spectra. The true values (x-axis) are shown against the recovered values with 1$\sigma$ uncertainties (y-axis). One-to-one lines are included in each panel. The reduced $\chi^2$ and average 1$\sigma$ uncertainties are provided in the lower right corner for each of the parameters. Overall, the recovery is accurate. However, panels with light-gray data points are deemed unreliable and were thus omitted from discussion in the main text.}
    \label{fig:sim}
\end{figure*}

For our recovery tests, we simulate 30 mock LEGA-C spectra, each with randomly drawn velocity dispersions, stellar ages, and elemental abundances. These values are drawn from a normal distributions with mean and spread that reflect the individual LEGA-C results. Next, we generate the 30 \texttt{alf} models. We then match each model with a LEGA-C spectrum that has the closest stellar age and metallicity. In order to add the appropriate noise to each model, we identify the LEGA-C spectrum that has the closest continuum shape to the \texttt{alf} model. We then regrid the model onto the same wavelength grid and use a high-order polynomial to match the model continuum shape to the LEGA-C spectrum. Finally, we take the S/N of the LEGA-C spectrum (typically $\approx60\,$\AA$^{-1}$), inflate the noise using the \texttt{alf} jitter term, and add in this random Gaussian noise to the \texttt{alf} model. The simulated error spectrum is computed by scaling down the simulated spectrum according to the S/N of the model.

Second, we fit the simulated spectra with \texttt{alf} using the identical setup to what is described in Section~\ref{sec:alf}. The results are shown in Figure~\ref{fig:sim} where we compare the fits to the true simulated values. The reduced $\chi^2$ and average 1$\sigma$ uncertainties for the 30 spectra are listed in each panel, and we also include one-to-one lines. For the most part, the fits are in good agreement with the simulated values.

Finally, we select which parameters can be constrained. The primary selection criterion is that the uncertainties on the fitted parameters must be $<0.15\,$dex. We make this selection to ensure that the PDFs for the elemental abundances do not hit the upper or lower prior limits. Alternatively we could extend the prior limits. However, the large uncertainties on some of the abundance measurements would require very large prior limits, and include regions that the models do not cover. Thus, the results would be based off extrapolated models. The second selection criterion is that the reduced $\chi^2<2.0$. This selection ensures the parameters are well-constrained to within error. 

The constrained parameters (velocity dispersion, stellar ages, jitter term, and 11 abundances) based on the average uncertainties and reduced $\chi^2$ are highlighted in Figure~\ref{fig:sim} in black. The parameters that do not meet our criteria (O, Na, K, Mn, Cu, Sr, Ba, and Eu) are shown in light gray. It is unsurprising that these parameters do not meet our criteria, as their absorption features are either weak or at redder wavelengths than targeted by the LEGA-C spectra.

\section{Stacking Versus Hierarchical Bayes}
\label{app:sim_stack}

In this section we use the simulated spectra described in Appendix~\ref{app:simulation} to test different spectral stacking methods. The results of these methods are compared with the results from hierarchical Bayesian modeling.

\begin{deluxetable}{cccccc}[b!]
\tablehead{\colhead{Method} & \colhead{Continuum} & \colhead{Smoothed} & \colhead{Weighing} & \colhead{Combination} & \colhead{Error}\\ \colhead{} & \colhead{Normalization} & \colhead{(Y/N)} & \colhead{} & \colhead{} & \colhead{Propagation}}
\tablecaption{ Stacking Methods 
\label{tab:stacking_methods}}
\startdata
stack1 & Median & N & None & Median & MAD \\
stack2 & Polynomial & N & None & Median & MAD \\
stack3 & Polynomial & Y & None & Mean & bootstrap \\
\enddata
\tablecomments{The three stacking methods tested in Appendix \ref{app:sim_stack} with results shown in Figure~\ref{fig:sim_stack}.}
\end{deluxetable}

\begin{figure*}
    \centering
    \includegraphics[width=\textwidth]{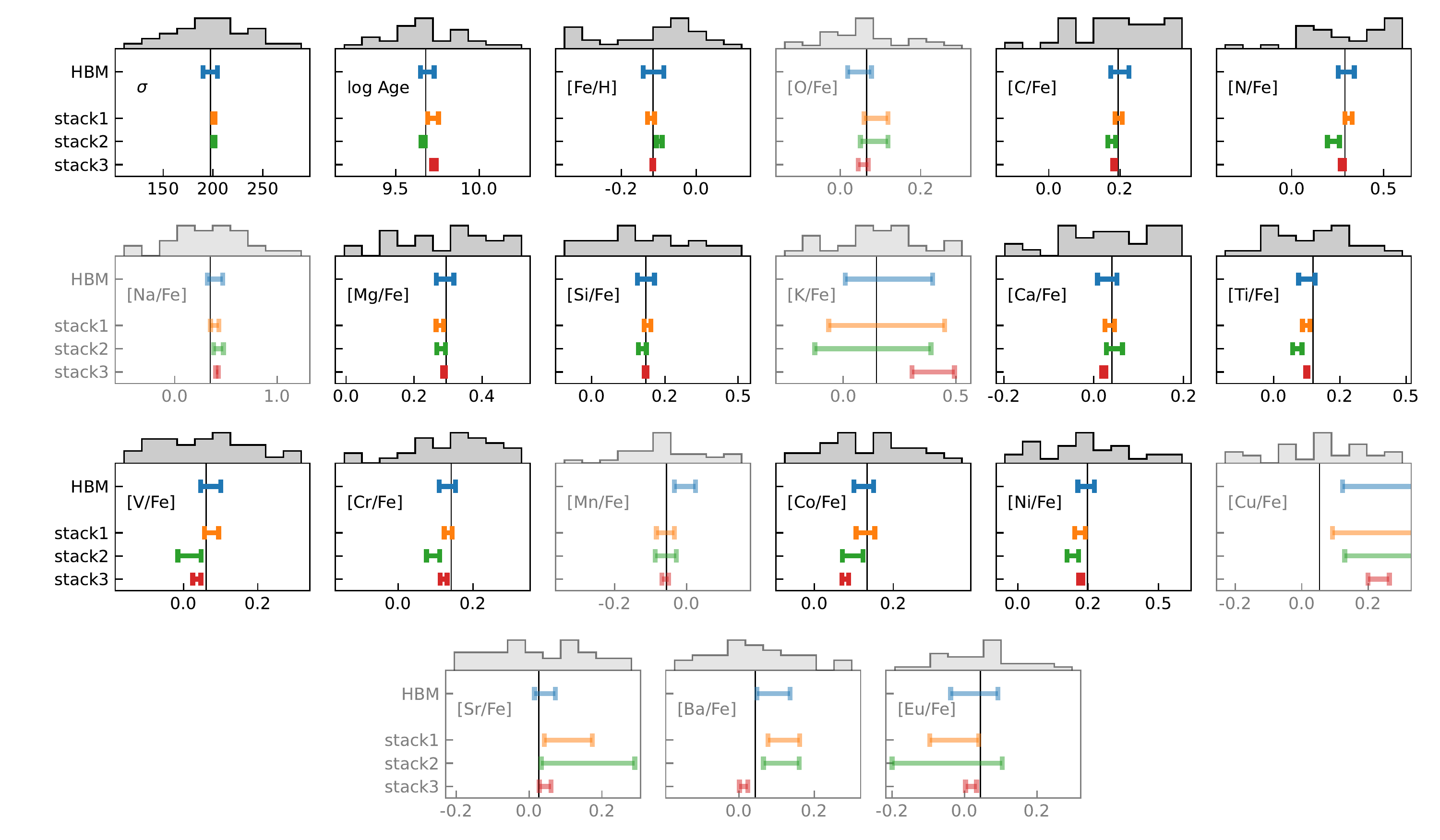}
    \caption{The 30 simulated spectra from Appendix A are combined using hierarchical Bayes and three different stacking methods to test how well the mean value of each parameter can be recovered. On top of each panel we show the distribution of simulated values. The vertical line is the value we try to recover (the mean of the distribution of parameters). The labeling of the different stacking methods correspond to Table 1.  The error bars represent the 16th and 84th percentiles on the PDF for each parameter. Parameters with low-transparency panels were deemed unreliable in Appendix~\ref{app:simulation}, and are thus omitted from discussion in the main text. All four methods produce accurate results, however HBM is the most consistent, with 1$\sigma$ uncertainties always encompassing the true value.}
    \label{fig:sim_stack}
\end{figure*}

Historically, when spectra have insufficient S/N, it is commonplace to take the spectra of many similar galaxies and stack them together. The act of stacking boosts the S/N and therefore enables us to derive average measurements. As described in the main text, the primary shortcoming of stacking is that it introduces systematic uncertainties and potential bias. For example, in order to stack the spectra, it is common to continuum normalized by fitting a high-order polynomial, smooth each spectrum to a constant velocity dispersion, interpolate them onto a global wavelength grid, and finally combine all the spectra via a weighted sum. Each of these steps introduces a different type of systematic uncertainty; fitting a polynomial runs the risk of removing part of the absorption features by over-fitting, smoothing introduces correlated noise and reduces the information content of the spectrum, linear interpolation does a poor job at conserving the noise level, and the way you combine the spectra (e.g. mean, light-weighted mean, median) affects the relative weighing of the individual spectra. For these reasons we turn to hierarchical Bayesian modeling (see Section~\ref{sec:hbm}). Here, we use simulated spectra from Appendix~\ref{app:simulation}, for which we know the ground truth stellar population parameters, and we test whether the hierarchical Bayesian modeling method quantitatively outperforms various stacking methods.

The simulated spectra are stacked according to the three methods summarized in Table~\ref{tab:stacking_methods} and then fitted using \texttt{alf}. Some of the differences between the stacking methods include: whether and how to remove the continuum before fitting, whether or not to smooth the spectra to the same velocity dispersion, whether and how to weigh each spectrum when combining, how to combine (e.g., mean vs median), and finally, how to propagate the errors through to an stacked error spectrum. 

The results from the stacked spectra are compared to the hierarchical Bayesian modeling method in Figure~\ref{fig:sim_stack} for various stellar population parameters and elemental abundance ratios. On the top of each panel we include a histogram showing the distribution of the 30 simulated spectra from Appendix~\ref{app:simulation}. The mean of the distribution is marked in each panel. This is the value we wish to recover. On the y-axis we label each method, where the stack number corresponds to Table~\ref{tab:stacking_methods}. The 1$\sigma$ uncertainties for each parameter are shown. All four methods reproduce the desired results quite closely (typically within 0.1\,dex). This result is reassuring given the various stacking methods found in the literature. The hierarchical Bayesian modeling method, however, produces the most consistent results to within error. Furthermore, the PDF more accurately represents which parameters we can actually constrain. For example, there are no strong Na features in the blue LEGA-C spectra, and thus the uncertainties on [Na/Fe] reflect this. Stacking methods often underestimate the error spectrum and therefore underestimate the uncertainties on individual elemental abundances.

\section{Hierarchical Bayesian Modeling}
\label{app:HBM}

In this section we derive the hierarchical Bayesian modeling method and its application to combining spectra. In the specific case of this paper, we first fit all of the individual spectra and then combine the resulting PDFs using the hierarchical Bayesian framework. Following the hierarchical Bayesian prescription, we demonstrate how to combine the spectra in ``post-processing.''

In the hierarchical framework, where we use $\alpha$ to describe the population parameters $\theta$, Bayes' theorem can be written,

\begin{equation} \label{eq:1}
P(\theta, \alpha | X) = \frac{P(X | \alpha, \theta)\cdot P(\alpha)\cdot P(\theta)}{P(X)}
\end{equation}

\noindent Or, expanded using the chain rule,

\begin{equation} \label{eq:2}
P(\theta, \alpha | X) = \frac{P(X | \theta) \cdot P(\theta | \alpha)\cdot P(\alpha)\cdot P(\theta)}{P(X)}
\end{equation}

\noindent Next, we re-write Hierarchical Bayes' Theorem using problem-specific variables,

\begin{equation} \label{eq:3}
\begin{split}
P(\{X_n\},\theta | \{S_n\}) & = \frac{P(\{S_n\}|\theta,\{X_n\}) \cdot P(\theta) \cdot P(\{X_n\})}{P(\{S_n\})} \\
&= \frac{P(\{S_n\}|\{X_n\}) \cdot P(\{X_n\}|\theta) \cdot P(\theta) \cdot P(\{X_n\})}{P(\{S_n\})},
\end{split}
\end{equation}

\noindent where $X_n$ is the set of individual parameters describing the $n^{\rm th}$ galaxy, and $\{X_n\}$ is the set of these parameters for all $N$ galaxies. $\theta$ is the population parameters that describe the distribution of $\{X_n\}$, and $\{S_n\}$ is the set of spectra for all N galaxies. For readability, the brackets will be dropped for the remainder of this section. Next, we put this equation in integral form,

\begin{equation} \label{eq:4}
P(X_n,\theta | S_n) = \prod_{n=1}^{N}\int \frac{P(S_n|X_n)\,P(X_n|\theta)\,P(\theta)\,P(X_n)}{P(S_n)}\,dX_n.
\end{equation}

\noindent Taking a uniform prior on $\theta$ and $X_n$, we can remove $\frac{P(\theta)}{P(S_n)}$ from the integral and set $P(X_n)=1$,

\begin{equation} \label{eq:5}
P(X_n,\theta | S_n) = \frac{P(\theta)}{P(S_n)}\prod_{n=1}^{N}\int P(S_n|X_n)\,P(X_n|\theta)\,dX_n.
\end{equation}

\subsection{Hierarchical Bayes in post-processing}

In this paper, each of the individual spectra are first fit with \texttt{alf}. As a result, we already have MCMC chains for each spectrum. Instead of fitting for the individual parameters ($X_n$) and population parameters ($\theta$) simultaneously, the above definitions can be re-written to include the pre-computed $X_n$ PDFs. The set of MCMC chains that describe the $X_n$ PDFs for all $N$ galaxies can be defined mathematically as,

\begin{equation} \label{eq:6}
X_n^{(k)} = P(X_n|S_n)
= \frac{P(S_n|X_n)\,P(X_n)}{P(S_n)},
\end{equation}

Note that $P(S_n|X_n)$ is actually $P(S_n|X_n,\theta)$ but we assert that the spectra do not care about the population parameters $\theta$ (e.g. galaxies don't talk to each other) and therefore we remove the dependence on $\theta$. Solving for $P(S_n|X_n)$, setting the prior on the individual parameters $P(X_n)=1$, and substituting into Eq.~\ref{eq:5},

\begin{equation} \label{eq:7}
P(X_n,\theta | S_n) =  \frac{P(S_n)\,P(\theta)}{P(S_n)}\prod_{n=1}^{N}\int P(X_n|S_n)\,P(X_n|\theta)\,dX_n.
\end{equation}

The first term in the integral, $P(X_n|S_n)$, describes the MCMC chains from the individual fits (Eq.~\ref{eq:6}) and the second term, $P(X_n|\theta)$, is the population model. This integral is too complicated to evaluate analytically, and thus we rely on finite sampling via MCMC. Eq.~\ref{eq:7} can be approximated as a sum over the chains, $k$, following the prescription in \citet{hogg_data_2018},

\begin{equation} \label{eq:8}
\int f(x)\,p(x)\approx \frac{1}{K}\sum_{k=1}^K f(x) \approx \frac{1}{K}\sum_{k=1}^K e^{\log f(x)},
\end{equation}

\noindent with the right-hand-side of this equation being a simple way to avoid numerical errors. Comparing Eqs.~\ref{eq:7} and \ref{eq:8}, we can define $f(x) \equiv P(X_n|\theta)$ (i.e. the population model). Taking the log of Eq.~\ref{eq:8} and substituting Eq.~\ref{eq:7}, we arrive at

\begin{equation} 
\log P(X_n,\theta | S_n) = \log P(S_n) - P(S_n) + \log P(\theta) - \log K + \sum_{n=1}^N\log\left[\sum_{k=1}^K e^{\log P(X_n|\theta)}\right].
\end{equation}

The first and second terms are the ``evidence'' because we do not care about the relative probability scaling, they can be ignored. The third term, $\log P(\theta)$, is the prior on the population parameters, which in our case is uniform and thus a constant that can also be ignored. Therefore, to calculate the posteriors on the global parameters given the MCMC chains of individual fits, we must evaluate the following:

\begin{equation} \label{eq:10}
\log P(X_n,\theta | S_n) \propto \sum_{n=1}^N\log\left[\sum_{k=1}^K e^{\log P(X_n|\theta)}\right].
\end{equation}

In this study, we set the population model to $P(X_n|\theta)\sim \mathcal{N}(\mu_{\rm pop},\, \sigma_{\rm pop}^2)$, such that each of the elemental abundances and stellar population parameters in a given bin are distributed normally, with a mean $\mu_{\rm pop}$ and spread $\sigma_{\rm pop}$. Finally, we use MCMC to sample the $\mu_{\rm pop}$ and $\sigma_{\rm pop}$ PDFs, where the likelihood is given by Eq.~\ref{eq:10}. In words, the likelihood function is given by the following: for each galaxy, evaluate $\mathcal{N}(\mu_{\rm pop},\, \sigma_{\rm pop}^2)$ at the location of the pre-computed MCMC chains. Sum this over all $K$ chains and then over all $N$ galaxies.


\end{document}